\documentclass[a4paper,11pt]{article}
\usepackage{jheppub}

\pdfoutput=1
\usepackage{hyperref}
\usepackage{graphicx}
\usepackage[utf8]{inputenc}
\usepackage{xspace}
\usepackage{amsmath,amssymb}
\usepackage{bbm}
\usepackage{xcolor}
\usepackage{booktabs}
\usepackage{subcaption}
\usepackage{mathtools}
\usepackage[ruled,norelsize]{algorithm2e}
\usepackage{empheq}
\usepackage{cancel}
\usepackage{multirow}
\usepackage{url}

\newcommand{\abar}{\bar{\alpha}}

\newcounter{subalgo}

\DeclareFontFamily{U}{rcjhbltx}{}
\DeclareFontShape{U}{rcjhbltx}{m}{n}{<->rcjhbltx}{}
\newcommand{\mim}{{\cal C}}

\title{Higher-order non-global logarithms from jet calculus}

\author[a]{Andrea Banfi,}
\author[b]{Fr\'ed\'eric A. Dreyer,}
\author[c]{Pier Francesco Monni.}

\affiliation[a]{Department of Physics and Astronomy, University of
  Sussex, Sussex House, Brighton, BN1 9RH, UK}
\affiliation[b]{Rudolf Peierls Centre for Theoretical Physics,
  Clarendon Laboratory, Parks Road, Oxford OX1 3PU, UK}
\affiliation[c]{CERN, Theoretical Physics Department, CH-1211 Geneva 23, Switzerland}

\preprint{CERN-TH-2021-157, OUTP-21-24P}
\abstract{ 
  Non-global QCD observables are characterised by a sensitivity to the
  full angular distribution of soft radiation emitted coherently in
  hard scattering processes. This complexity poses a challenge to
  their all-order resummation, that was formulated at the
  leading-logarithmic order about two decades ago.
  In this article we present a solution to the long-standing problem
  of their resummation beyond this order, and carry out the first
  complete next-to-leading logarithmic calculation for non-global
  observables. This is achieved by solving numerically the recently
  derived set of non-linear differential equations which describe the
  evolution of soft radiation in the planar, large-$N_c$ limit.
  As a case study we address the calculation of the transverse energy
  distribution in the interjet rapidity region in $e^+e^-\to$ dijet
  production.
  The calculation is performed by means of an algorithm that we
  formulate in the language of jet-calculus generating functionals,
  which also makes the resummation technique applicable to more
  general non-global problems, such as those that arise in hadronic
  collisions.
  We find that NLL corrections are substantial and their inclusion
  leads to a significant reduction of the perturbative scale
  uncertainties for these observables.
  The computer code used in the calculations is made publicly
  available.}

\begin{document}
\setlength{\parskip}{0pt}
\maketitle
\flushbottom

\section{Introduction}
\label{sec:intro}
The topic of non-global QCD observables~\cite{Dasgupta:2001sh} has
received significant attention in recent years.
A first reason is due to their theoretical complexity. These
observables are characterised by kinematic constraints on limited
angular regions of the radiation phase space, which leads to a rich
structure in perturbation theory.
This class of logarithmic corrections to physical observables was
discovered and resummed at leading logarithmic (LL) order in the
planar, large-number-of-colour (large-$N_c$) limit about 20 years
ago~\cite{Dasgupta:2001sh,Dasgupta:2002bw,Banfi:2002hw}, and methods
to calculate finite-$N_c$ effects are also well
established~\cite{Hatta:2013iba,Hagiwara:2015bia,Hatta:2020wre}.
While subleading-colour corrections are commonly numerically small in
known applications, their study is of importance for the understanding
of the structure of super-leading logarithmic corrections in
non-global observables at hadron
colliders~\cite{Forshaw:2006fk,Forshaw:2008cq,Becher:2021zkk}.

The calculation of non-global corrections to higher orders largely
remains an open problem.
Several new formulations of the resummation have been proposed in
recent
years~\cite{Becher:2015hka,Becher:2016mmh,Caron-Huot:2015bja,Larkoski:2015zka,Banfi:2021owj},
accompanying a large amount of applications to perturbative
calculations of collider observables at LL, and recently also
including some class of NLL corrections (see
e.g.~\cite{Forshaw:2009fz,Rubin:2010fc,Banfi:2010pa,DuranDelgado:2011tp,Dasgupta:2012hg,Schwartz:2014wha,Becher:2015hka,Becher:2016omr,Neill:2016stq,Caron-Huot:2016tzz,Larkoski:2016zzc,Becher:2017nof,Martinez:2018ffw,Balsiger:2018ezi,Neill:2018yet,Neill:2018yet,Balsiger:2019tne,Balsiger:2020ogy,Ziani:2021dxr}).
Their study is also motivated by theoretical interests related to the
connection between their dynamics and the high-energy limit of
scattering
amplitudes~\cite{Weigert:2003mm,Hatta:2008st,Caron-Huot:2015bja}.

A second reason why non-global observables are interesting is their
ubiquitous occurence at colliders, for instance via the use of jets or
often when specific fiducial cuts are applied in experimental
measurements. In the context of the precision physics programme of
present and future particle colliders, it is therefore paramount to
gain theoretical control over non-global corrections to physical
observables.

Finally, an understanding of their dynamics is instrumental in the
context of developing more accurate parton-shower algorithms (see
e.g.\
Refs.~\cite{Dasgupta:2018nvj,Bewick:2019rbu,Dasgupta:2020fwr,Forshaw:2020wrq,Platzer:2020lbr,Hamilton:2020rcu,Nagy:2020rmk,Nagy:2020dvz,Karlberg:2021kwr,Dulat:2018vuy,Gellersen:2021eci,Hamilton:2021dyz}). Specifically,
the resummation of next-to-leading logarithmic (NLL) non-global
logarithms is a crucial ingredient for the development of NNLL
algorithms, that are necessary to achieve sufficiently accurate event
simulation both at present and future colliders.

In a recent article~\cite{Banfi:2021owj} we have developed a framework
for the resummation of non-global observables in the planar limit,
which relies on a set of non-linear evolution equations that describe
the dynamics of soft radiation at different energy
scales. Ref.~\cite{Banfi:2021owj} also demonstrates the correctness of
the framework by comparing a calculation at fixed perturbative order
in this formalism to the full QCD result for the energy and transverse
energy distributions in the rapidity gap between two cone jets
produced in electron-positron annihilation.

In this article, we instead address the solution of the equations of
Ref.~\cite{Banfi:2021owj} at all perturbative orders, hereby achieving
a first complete NLL resummation for a non-global observable in the
large-$N_c$ limit.
The paper is structured as follows. Sec.~\ref{sec:recall} contains a
summary of the findings of Ref.~\cite{Banfi:2021owj}.
Sec.~\ref{sec:functionals} reformulates the evolution equations in
terms of a well known method used in jet calculus, that of generating
functionals~\cite{Konishi:1979cb,Bassetto:1984ik,Dokshitzer:1991wu}.
This formulation has two important advantages. Firstly, it provides us
with a probabilistic picture to solve the evolution equations using a
Markov chain Monte-Carlo algorithm. Secondly, the resulting algorithm
describing the evolution of the soft radiation in the planar limit is
\textit{independent} of the underlying hard scattering process, and
therefore can be readily applied to other observables and reactions,
such as jet production at hadron colliders.
The algorithm is given in detail in Sec.~\ref{sec:PT-solution}, which
is arguably the most technical part of the article. 
The reader uninterested in the technical aspects of the calculation
can skip directly from Sec.~\ref{sec:functionals} to
Sec.~\ref{sec:results} where the numerical results are presented.
In Sec.~\ref{sec:results} we report NLL predictions for the transverse
energy distribution in the interjet rapidity slice in $e^+e^-\to \,2$
jets, and discuss the impact of NLL corrections as well as the
reduction in the perturbative uncertainty.
Finally, Sec.~\ref{sec:conclusions} contains our conclusions.

\section{Resummation for $E_t$ in the interjet rapidity gap}
\label{sec:recall}
\begin{figure}[htbp]
  \centering
  \includegraphics{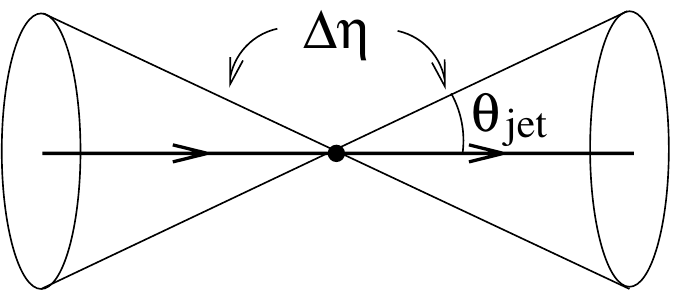}
  \caption{The rapidity slice where the measurement is performed.}
  \label{fig:obs}
\end{figure}

As in ref.~\cite{Banfi:2021owj}, we consider the production of two
jets in $e^+e^-$ annihilation at a centre-of-mass energy
$\sqrt{s}$. We study a non-global observable defined by measuring
hadrons in a rapidity slice between the two jets. This is defined as
the rapidity region between two cones of opening angle
$\theta_{\rm jet}$ around the thrust axis (see
Fig.~\ref{fig:obs}). The width of such a rapidity slice is
\begin{equation}
\label{eq:c_def}
\Delta\eta \coloneqq \ln\frac{1+c}{1-c} \,,\qquad c=\cos \theta_{\rm
  jet}\,.
\end{equation}
Examples of non-global observables are those studied in
ref.~\cite{Banfi:2021owj}, namely the total energy or transverse
energy of hadrons inside the slice. In the following we consider the
latter for the sake of arguments, although the formalism introduced in
this section also applies to the calculation of the energy
distribution.

The cumulative distribution for this observable to be less than
$v\equiv E_t$ is defined as follows
\begin{equation}
\label{eq:sigma-def}
\Sigma(v)\coloneqq\frac{1}{\sigma_0}\int_{0}^v \frac{d\sigma}{d v^\prime} d v^\prime\,,
\end{equation}
where $\sigma_0$ is the Born cross section for $e^+e^-\to$ hadrons.
Without any emissions, at the lowest order in perturbation theory,
$\Sigma(v)=1$, and the event is made up of a quark of momentum $p_1$
and an antiquark of momentum $p_2$, back-to-back and aligned along the
thrust axis. When extra radiation is considered, $\Sigma(v)$ can be
expressed as~\cite{Banfi:2021owj}
\begin{equation}
\label{eq:master}
\Sigma(v) \coloneqq \sum_{n=2}^{\infty}{\cal H}_n\otimes S_n(v) = {\cal H}_2\otimes S_2(v) + {\cal H}_3\otimes S_3(v) +\cdots
\end{equation}
where the hard factors 
\begin{equation}
{\cal H}_n\coloneqq {\cal H}_{1\dots n}
\end{equation}
describe configurations with $n$ hard QCD partons along the light-like
directions $n_1,\dots,n_n$ (with $n^2_i=0$ and $|\vec{n}|^2=1$), while
the soft factors 
\begin{equation}
S_n \coloneqq S_{1\dots n}
\end{equation}
describe the emission of soft radiation off a hard system with $n$
hard emitters along the same directions. The convolutions in
Eq.~\eqref{eq:master} are meant to indicate that the directions of the
hard emitters in the hard and soft factors are the same, namely
\begin{equation}
\label{eq:convolution}
{\cal H}_n\otimes S_n(v) \coloneqq \int\left(\prod_{i=i}^n d^{2}\Omega_i\right)
{\cal H}_{1\dots n} \times S_{1\dots n} (v),
\end{equation}
where $\Omega_i$ indicates the solid angle of the $i$-th hard emitter,
namely the direction of the $\vec{n}_i$ vector, specified by a
longitudinal ($\theta$) and an azimuthal ($\phi$) angle. In writing
Eq.~\eqref{eq:convolution} we assumed that all IRC divergences cancel
in the definition of ${\cal H}_n$ and $S_n$, and the four-dimensional
limit can be taken in the angular integration~\cite{Banfi:2021owj}.
Each of the above ingredients admits a perturbative expansion in the
strong coupling constant~\cite{Banfi:2021owj}
\begin{align}
\label{eq:conventions}
 {\cal H}_n = \sum_{i=n-2}^\infty\frac{\alpha_s^i}{(2\pi)^i}  {\cal
  H}_n^{(i)},\qquad
 S_n = \sum_{i=0}^\infty\frac{\alpha_s^i}{(2\pi)^i} S_n^{(i)},
\end{align}
where the normalisation is defined such that at the Born level one has
\begin{equation}
{\cal H}_2^{(0)}=\delta^{(2)}(\Omega_1-\Omega_q)\delta^{(2)}(\Omega_2-\Omega_{\bar
             q})\coloneqq \delta(\cos\theta_1-1) \delta(\cos\theta_2+1)
\delta(\phi_1) \delta(\phi_2)\,,\quad S_n^{(0)}=1\,.
\end{equation}
At LL, the cumulative distribution is given by the convolution of the
LO hard factor ${\cal H}_2$ with the LL soft factor $S_2$.
At NLL, one needs to include the LO hard factor ${\cal H}_3$
convoluted with the LL soft factor $S_3$, as well as the NLO hard
factor ${\cal H}_2$ convoluted with the NLL soft factor $S_2$.

The observables considered in ref.~\cite{Banfi:2021owj} were all
additive, which implies that the observable constraint on the soft
emissions contributing to $S_n$ factorises under the Laplace transform
\begin{equation}
\label{eq:additive-Laplace}
\Theta\left(v - \sum_i v(k_i)\Theta_{\rm in}(k_i)\right) 
= \frac{1}{2\pi i}
\int_{\gamma}\frac{d\nu}{\nu} e^{\nu v} \prod_i u(k_i)
\,,
\end{equation}
where the trigger function $\Theta_{\rm in}(k)$ is 1 if the particle
$k$ is inside the measurement region, and zero otherwise, and the
contour $\gamma$ lies parallel to the imaginary axis and to the right
of all singularities of the integrand. The quantity $u(k)$ is the
following ``source'' function:
\begin{equation}
\label{eq:source}
u(k) = \Theta_{\rm out}(k) + \Theta_{\rm in}(k) e^{-\nu v(k) } \,,
\end{equation}
with $\Theta_{\rm out}(k)=1-\Theta_{\rm in}(k)$.
Following ref.~\cite{Banfi:2021owj}, in the large-$N_c$ limit we can
then define the Laplace transform of the soft factors entering the NLL
calculation
\begin{equation}
  \label{eq:S2-S3-additive-largeNc}
  \begin{split}
  S_2(v)  =  \frac{1}{2\pi i}
\int_{\gamma}\frac{d\nu}{\nu} e^{\nu v} G_{12}[Q;u]\,,\quad
  S_3(v)  =  \frac{1}{2\pi i}
\int_{\gamma}\frac{d\nu}{\nu} e^{\nu v} G_{13}[Q;u]G_{23}[Q;u]\,.
  \end{split}
\end{equation}
The evolution of the $G_{ij}[Q;u]$ functionals is governed by the
differential equation (in the following we set $\{ij\}=\{12\}$ unless
otherwise specified, but the same considerations hold for the
evolution of a generic dipole $\{ij\}$)
\begin{align}
\label{eq:NLL-evolution-kt-diff}
  Q \partial_Q G_{12}[Q;u] ={\mathbb K}[G[Q;u],u]\,.
\end{align}
The evolution kernels are derived in Ref.~\cite{Banfi:2021owj}, and we
report them in $4-2\epsilon$ dimensions below. The LL kernel, relevant
for the evolution of $S_3$ reads
\begin{align}
\label{eq:LLkernel}
{\mathbb K}^{\rm LL}&[G[Q;u],u] \coloneqq \int [d k_a] \abar(Q) w^{(0)}_{12}(k_a)\\
&\hspace{1cm}\times\left(G_{1a}[Q;u]
  G_{a2}[Q;u] u(k_a) - G_{12}[Q;u]\right)Q\delta(Q-k_{ta})\notag\,.
\end{align}
The tree-level eikonal squared amplitude is defined as
\begin{equation}
\label{eq:eik0}
w^{(0)}_{ij}(k) = 8\pi^2\frac{\mu^{2\epsilon}}{k_t^2} \,,\quad k_t^2 \coloneqq (k_t^{(ij)})^2=
2\frac{(k_i\cdot k)(k\cdot k_j)}{(k_i\cdot k_j)}
\,,
\end{equation}
where $k_t$ denotes the transverse momentum of emission $k$ w.r.t. the
emitting $\{ij\}$ dipole (note that in the equation above
$k_{i,j}\equiv p_{i,j}$ if it refers to one of the hard legs).
In the same variables, the corresponding phase space measure in
$4-2\epsilon$ dimensions reads
\begin{equation}
\label{eq:phase-space-kt}
[d k]\coloneqq \frac{d \eta}{2}\frac{d^{2-2\epsilon}k_{t}}{(2\pi)^{3-2\epsilon}}\,,
\end{equation}
where the rapidity bound in the soft limit is given by
\begin{equation}
\label{eq:rapidity-bound}
|\eta| \lesssim \ln
\frac{\sqrt{2 \,k_i\cdot k_j}}{k_t}\,.
\end{equation}

A discussion about the choice of ordering variable and its relation to
symmetries of the multi-particle squared amplitude is reported in
Appendix~\ref{sec:symmetries}.  In Eq.~\eqref{eq:LLkernel} we also
defined $\abar= N_c \alpha_s(\mu)/\pi$, where $\alpha_s$ is the QCD
coupling in the $\overline{\rm MS}$ scheme, satisfying the
renormalisation group equation (RGE)
\begin{equation}
\label{eq:RGEplanar}
\frac{d\abar(\mu)}{d\ln\mu^2} = -\bar{\beta}(\abar) = -\abar
\left(\bar{\beta}_0 \abar + \bar{\beta}_1 \abar^2 + \dots\right)\,,
\end{equation}
where $\bar{\beta}_i$ are obtained from the large-$N_c$ limit of the
coefficients of the QCD beta function as
\begin{equation}
\bar{\beta}_i = \lim_{N_c\to\infty} \left(\frac{\pi}{N_c}\right)^{i+1}\beta_i\,.
\end{equation}
We work in the normalisation in which $\beta_0 = (11 C_A-2
n_F)/(12\pi)$ and $\bar{\beta}_0=11/12$.
The coupling in the LL kernel~\eqref{eq:LLkernel} evolves at one loop
(i.e.\ setting $\bar{\beta}_1=0$ in Eq.~\eqref{eq:RGEplanar}), while it
evolves at two loops in the NLL kernel defined in the following.
The NLL kernel which governs the evolution of $S_2$ reads
\begin{align}
\label{eq:kernel_NLL}
{\mathbb K}^{\rm NLL}[G[Q;u],u]\coloneqq {\mathbb K}^{\rm RV+VV}[G[Q;u],u]+{\mathbb K}^{\rm RR}[G[Q;u],u]-{\mathbb K}^{\rm DC}[G[Q;u],u]\,.
\end{align}
The three contributions to the above equation describe three sources
of NLL corrections.  The kernel correction due to the subtracted
virtual and real-virtual corrections reads
\begin{align}
\label{eq:Uvirtual}
{\mathbb K}^{\rm RV+VV}&[G[Q;u],u] \coloneqq \int [d k_a] \abar(Q) w^{(0)}_{12}(k_a)
\,\bigg(1 +\abar(Q)\,\bar{K}^{(1)}\bigg)\\
&\hspace{1cm}\times\left(G_{1a}[Q;u]
  G_{a2}[Q;u] u(k_a) - G_{12}[Q;u]\right)Q\delta(Q-k_{ta})\notag\,,
\end{align}
where $\bar{K}^{(1)}$ is obtained from the two-loop cusp anomalous
dimension in the large-$N_c$ limit
\begin{equation}
\label{eq:cusp}
\bar{K}^{(1)} = \lim_{N_c\to\infty} \frac{2}{N_c}\,K^{(1)} = \frac{67}{36}-\frac{\pi^2}{12}\,.
\end{equation}
Then, the double real corrections are given by
\begin{align}
\label{eq:Ureal}
{\mathbb K}^{\rm RR}&[G[Q;u],u] \coloneqq  \int [d k_{a}]\int
                      [d k_{b}] \,\abar^2(Q)Q\delta(Q-k_{t(ab)}) {\Theta(k_{ta}-k_{tb}^\prime)}\\
&\times\left[ \bar{w}^{(gg)}_{12}(k_b,k_a)  G_{1b}[Q;u]
  G_{ba}[Q;u]G_{a2}[Q;u]u(k_a)u(k_b)\right. \notag\\
&\left.\hspace{1cm}+ \bar{w}^{(gg)}_{12}(k_a,k_b) G_{1a}[Q;u]
  G_{ab}[Q;u]G_{b2}[Q;u]u(k_a)u(k_b)\right.\notag\\
&\left. \hspace{2cm} - \left(\bar{w}^{(gg)}_{12}(k_b,k_a) +\bar{w}^{(gg)}_{12}(k_a,k_b)\right)G_{1(ab)}[Q;u]
  G_{(ab)2}[Q;u]u(k_{(ab)})\right]\notag\,,
\end{align}
where $k_{tb}^\prime$ denotes the transverse momentum of $k_b$ with
respect of the $\{12\}$ dipole, namely
\begin{equation}
(k_{tb}^\prime)^2 =
2\frac{(p_1\cdot k_b)(k_b\cdot p_2)}{(p_1\cdot p_2)}\,.
\end{equation}
To properly define $\bar{w}^{(gg)}_{12}(k_a,k_b)$ we need first to
introduce the colour-ordered double soft squared amplitude at tree
level~\cite{Campbell:1997hg,GehrmannDeRidder:2005cm}\footnote{We
  thank Keith Hamilton for an independent derivation of these squared
  amplitudes.}
\begin{align}
  \label{eq:tildew}
\tilde{w}^{(0)}_{12}(k_a,k_b)&=2\,(2\pi)^4\mu^{4\epsilon}\bigg[\frac{s_{12}^2}{s_{1a}s_{ab2}s_{1ab}s_{b2}}
  +
  \frac{1-\epsilon}{s_{ab}^2}\left(\frac{s_{1a}}{s_{1ab}}+\frac{s_{b2}}{s_{ab2}}-1\right)^2\notag\\
&+\frac{s_{12}}{s_{ab}}\left(\frac{1}{s_{1a}s_{b2}}+\frac{1}{s_{1a}s_{ab2}}+\frac{1}{s_{b2}s_{1ab}}-\frac{4}{s_{1ab}
  s_{ab2}}\right)\bigg]\,,
\end{align}
where the Lorentz invariants $s_{i\dots k}$ indicate the standard
Mandelstam variables.
Following~\cite{Banfi:2021owj}, we define
$\bar{w}^{(gg)}_{12}(k_a,k_b)$ as the correlated contribution to
$\tilde{w}^{(0)}_{12}(k_a,k_b) $, as follows
\begin{align}
\label{eq:double-soft-real}
 \bar{w}^{(gg)}_{12}(k_a,k_b) \coloneqq\tilde{w}^{(0)}_{12}(k_a,k_b)-\frac{1}{2}w^{(0)}_{12}(k_a)w^{(0)}_{12}(k_b) \,,
\end{align}
where $\frac 12 w^{(0)}_{12}(k_a)w^{(0)}_{12}(k_b)$ represents the independent emission contribution.\footnote{Note, however, that the separation of the independent
  contribution is immaterial at the level of the single colour flow,
  and only makes physical sense at the level of the sum
  $\tilde{w}^{(0)}_{12}(k_a,k_b) + \tilde{w}^{(0)}_{12}(k_b,k_a)$.}
The counter-term in the r.h.s. of Eq.~\eqref{eq:Ureal} is built upon
the massless momentum $k_{(ab)}$ defined by the following kinematic
map
\begin{equation}
\label{eq:map_ab}
{\mathbb P} : \{k_a,k_b\}\rightarrow k_{(ab)} = 
{\left(k_{t(ab)}\cosh \eta_{(ab)},\vec{k}_{t(ab)},k_{t(ab)}\sinh \eta_{(ab)}\right)}
\,,
\end{equation}
where $k_{t(ab)}$ and $\eta_{(ab)}$ denote the transverse momentum and
rapidity of $k_a+k_b$ in the $\{12\}$ dipole rest
frame. Eq.~\eqref{eq:map_ab} is expressed in the $\{12\}$ dipole rest
frame where the $\{12\}$ dipole is aligned with the $z$ axis. A
Lorentz transformation (a rotation followed by a boost) must be then
applied to $k_{(ab)}$ to express it in the event frame.
Last, we \textit{subtract} the iteration of the LL kernel
\begin{align}
\label{eq:Udc}
{\mathbb K}^{\rm DC}&[G[Q;u],u] \coloneqq \int [d k_a] \int [d k_b] \abar^2(Q) Q\delta(Q-k_{ta}) \Theta(k_{ta}-k_{tb})\\
&\times\biggl[w^{(0)}_{12}(k_a) \left(w^{(0)}_{1a}(k_b) -\frac{1}{2}
  w^{(0)}_{12}(k_b)\right)  G_{1b}[Q;u]
  G_{ba}[Q;u]G_{a2}[Q;u]u(k_a)u(k_b) \notag\\
&\hspace{1cm}+ w^{(0)}_{12}(k_a) \left(w^{(0)}_{a2}(k_b)
  -\frac{1}{2} w^{(0)}_{12}(k_b)\right) G_{1a}[Q;u]
  G_{ab}[Q;u]G_{b2}[Q;u] u(k_a)u(k_b)\notag\\
&\hspace{2cm}-w^{(0)}_{12}(k_a) \left(w^{(0)}_{1a}(k_b)  +w^{(0)}_{a2}(k_b)- w^{(0)}_{12}(k_b)\right) G_{1a}[Q;u]
  G_{a2}[Q;u]u(k_a)\biggr]\notag\,.
\end{align}
In the above expression, with a little abuse of notation, we denoted
with $k_{tb}$ the transverse momentum of $k_b$ with respect to the
\textit{emitting} dipole, that is each term should be interpreted as
follows
\begin{equation}
    \label{eq:ktij-def}
    w^{(0)}_{ij}(k_b) \Theta(k_{ta}-k_{tb}) \coloneqq w^{(0)}_{ij}(k_b) \Theta(k_{ta}-k^{(ij)}_{tb})\,,
  \end{equation}
  where $k^{(ij)}_{tb}$ is the transverse momentum of $k_b$ with
  respect to the ``emitting'' dipole $\{ij\}$ (see also
  Eq.~\eqref{eq:eik0})
\begin{equation}
(k^{(ij)}_{tb})^2 =
2\frac{(k_i\cdot k_b)(k_b\cdot k_j)}{(k_i\cdot k_j)}\,,
\end{equation}
with $k_{i,j}\equiv p_{i,j}$ if it refers to one of the hard legs.
The quantity $G_{ij}$ satisfies the boundary condition
\begin{equation}
\label{eq:initial-cond}
G_{ij}[Q;u] = 1~{\rm for}~ Q = 0\,,
\end{equation}
and the normalisation $G_{ij}[Q;1]=1$.
In taking the four-dimensional limit of the above equations some care
is required since the boundary condition has to be deformed by
introducing an appropriate non-perturbative prescription. In this
article we consider implementing the following
procedure~\cite{Banfi:2021owj}
\begin{equation}
\label{eq:initial-cond-4D-alt}
\alpha_s(k) = \alpha_s(Q_0) = 0,\qquad k \leq Q_0 \,,
\end{equation}
where $Q_0$ is defined below (see Eq.~\eqref{eq:Q0def}) and it is
above the Landau singularity.
This simply amounts to modifying the boundary
condition~\eqref{eq:initial-cond} such that $G_{ij}[Q;u] = 1$ for
$Q \leq Q_0$.
We also stress that in four dimensions the collinear singularity
$k_a\parallel k_b$ in the r.h.s. of Eq.~\eqref{eq:Udc} is regulated by
the requirement that at NLL the two soft gluons $k_a$ and $k_b$ cannot
be inside the slice simultaneously as this configuration would in the
end produce only a NNLL correction.

A comment is in order about the applicability criteria of the
evolution equations given in this section. As presented,
Eq.~\eqref{eq:NLL-evolution-kt-diff} can be used for the resummation
of NLL corrections to non-global observables which do not exhibit
logarithmic sensitivity to configurations in which the soft gluons are
radiated collinear to the emitting leg, which translates into the
absence of Sudakov double logarithms in their perturbative
expansion. Such observables are purely single logarithmic, i.e.\ the
dominant tower of logarithmic corrections are of the form
$\alpha_s^n L^n$.
Correspondingly, the precise form of the upper kinematic bound on the
emission's rapidity~\eqref{eq:rapidity-bound} is irrelevant in the
resummation of such observables, and it can be relaxed and replaced by
a finite (albeit sufficiently large) rapidity buffer for the radiation
within each emitting dipole.

For observables with a double logarithmic perturbative expansion, the
resummation presented here must be supplemented with the correct
resummation of the corresponding collinear logarithms (obtained with
standard techniques for global observables), and the double counting
between the two regions (i.e.\ the soft \text{and} collinear limit)
must be consistently subtracted. 
We do not address this subtraction in the present article.

\section{Integral equations and the generating functional method}
\label{sec:functionals}
In this article we wish to formulate a solution to the above
integro-differential equations in terms of an algorithmic
procedure. Moreover, while the above equations have been derived for
the family of additive observables, the dynamics they describe is
completely general and governs the resummation of non-global QCD
corrections in more generic cases. As a first step, we therefore wish
to re-write the evolution equations using a language that makes them
suitable for: \textit{i)} a numerical implementation via a Monte-Carlo
algorithm; \textit{ii)} the application to a generic non-global
observable sensitive to soft radiation at large angles.

To carry out this extension, we reinterpret the evolution equations
given in the previous section by exploiting a theoretical tool that
has been widely used in the area of jet calculus, the generating
functional method~\cite{Konishi:1979cb,Bassetto:1984ik,Dokshitzer:1991wu}.

We reinterpret the source $u(k)$ as a \textit{probing function}, whose
role is to assign a tag to a real emission $k$. The probability
associated with a state of $n$ real partons $d P_n^{\{12\}}$ produced
within a dipole $\{12\}$ is then defined by the following functional
derivative
\begin{equation}
\label{eq:emprob}
dP_n^{\{12\}} = \left. \mim (n)\left(\prod_{i=1}^n[d k_i]\frac{\delta}{\delta
    u(k_i)}\right) Z_{12}[Q;\{u\}] \right|_{\{u\}=0}\,,
\end{equation}
where the action of the functional derivative on $Z_{12}[Q;\{u\}]$ is
defined by
\begin{equation}
\frac{\delta}{\delta
    u(k_i)} u(k) \coloneqq \bar{\delta}(k-k_i) = 2 (2\pi)^{3-2\epsilon}\delta^{(2-2\epsilon)}(\vec{k}_t-\vec{k}_{ti})\delta(\eta-\eta_{i}) ,
\end{equation}
with the transverse momentum $\vec{k}_{ti}$ and rapidity $\eta_{i}$
being defined w.r.t.\ the emitting colour dipole.
The quantity $\mim (n)$ is an appropriate combinatorial factor for a
state with $n$ (not necessarily identical) particles. For identical
particles one simply has $\mim(n) = 1/n!$. The above equation defines
the \textit{generating} functional $Z_{12}[Q;\{u\}]$, whose
probabilistic interpretation~\eqref{eq:emprob} is crucial to derive a
Monte-Carlo procedure for its calculation.

We can now reinterpret the factorisation of the NLL cumulative cross
section~\eqref{eq:master} for an observable $V(\{k_i\}) < v$ in terms
of generating functionals simply by summing over all possible
configurations
\begin{align}
\label{eq:master_genfun}
\Sigma(v) &= {\cal H}_2\otimes \left[\sum_{i=0}^{\infty}\int
            d P_i^{\{12\}} \Theta(v-V(\{k_i\}))\right] \notag\\
  & + {\cal H}_3\otimes \left[\left(\sum_{i=0}^{\infty}\int
            d P_i^{\{13\}} \right)\, \left(\sum_{j=0}^{\infty}\int
            d P_j^{\{23\}} \right)\Theta(v-V(\{k_i\}, \{k_j\}))\right]
    +{\cal O}({\rm NNLL})\,,
\end{align}
where the zero-th terms of the above sums are equal to one.
To calculate the above probabilities, we observe that the generating
functional $Z_{12}[Q;\{u\}]$ satisfies the same evolution equations as
the Laplace transform of the soft
factor~\eqref{eq:NLL-evolution-kt-diff} with the same boundary
conditions and with the source $u(k)$ now playing the role of the
probing function. In the case of the generating functional, it is more
convenient to recast these equations in integral form which, as we
will see shortly, offers a simple probabilistic interpretation that
can be exploited to construct a Monte Carlo procedure to calculate
their solution.
Following the derivation in Section 4 of Ref.~\cite{Banfi:2021owj}, we
introduce the NLL Sudakov form factor associated with the no-emission
probability within the dipole $\{12\}$
\begin{equation}
\label{eq:sudakov_NLL_def}
\ln\Delta_{12}(Q) = - \int [d k] \Theta(Q-k_{t})
\abar(k_{t}) w^{(0)}_{12}(k) \left(1+\abar(k_t)
  \bar{K}^{(1)}\right)\,,
\end{equation}
and we take the $\ln Q$ derivative of $Z_{12}[Q;\{u\}]/\Delta_{12}(Q)$
\begin{equation}
Q\partial_Q \frac{Z_{12}[Q;\{u\}]}{\Delta_{12}(Q)} = \frac{{\mathbb K}[Z[Q;\{u\}],u]}{\Delta_{12}(Q)}-Z_{12}[Q;\{u\}]\frac{Q\partial_Q \Delta_{12}(Q)}{\Delta^2_{12}(Q)}\,.
\end{equation}
Using the NLL kernel~\eqref{eq:kernel_NLL}
and~\eqref{eq:sudakov_NLL_def} in the above equation, and integrating
over $\ln Q$ with the boundary condition~\eqref{eq:initial-cond} for
$Z_{12}$ leads to the integral equation~\cite{Banfi:2021owj}
\begin{align}
\label{eq:integral_NLL}
Z_{12}[Q;\{u\}] = {\mathbb K}_{\rm int}^{\rm RV+VV}[Z[Q;u],u] +
  {\mathbb K}_{\rm int}^{\rm RR}[Z[Q;u],u] - {\mathbb K}_{\rm int}^{\rm DC}[Z[Q;u],u]\,,
\end{align}
where we have defined the \textit{integrated} kernels as
\begin{align}
\label{eq:Uvirtualint}
 {\mathbb K}_{\rm int}^{\rm RV+VV}&[Z[Q;u],u] = \Delta_{12}(Q) +  \int [d k_a] \abar(k_{ta}) w^{(0)}_{12}(k_a)
                  \,\bigg(1 +\abar(k_{ta})\,\bar{K}^{(1)}\bigg)\frac{\Delta_{12}(Q)}{\Delta_{12}(k_{ta})}\notag\\
  &\hspace{0.5cm}\times Z_{1a}[k_{ta};\{u\}]
    Z_{a2}[k_{ta};\{u\}] u(k_a) \Theta(Q-k_{ta})\,,
\end{align}
\begin{align}
\label{eq:RRint}
 {\mathbb K}_{\rm int}^{\rm RR}&[Z[Q;u],u] =  \int [d k_{a}]\int
                      [d k_{b}] \,\abar^2(k_{t(ab)})\Theta(Q-k_{t(ab)}) {\Theta(k_{ta}-k_{tb}^\prime)}\frac{\Delta_{12}(Q)}{\Delta_{12}(k_{t(ab)})}\notag\\
&\hspace{0.5cm}\times\left[ \bar{w}^{(gg)}_{12}(k_b,k_a)  Z_{1b}[k_{t(ab)};\{u\}]
  Z_{ba}[k_{t(ab)};\{u\}]Z_{a2}[k_{t(ab)};\{u\}]u(k_a)u(k_b)\right. \notag\\
&\left.\hspace{0.5cm}+ \bar{w}^{(gg)}_{12}(k_a,k_b) Z_{1a}[k_{t(ab)};\{u\}]
  Z_{ab}[k_{t(ab)};\{u\}]Z_{b2}[k_{t(ab)};\{u\}]u(k_a)u(k_b)\right.\\
&\left. \hspace{0.5cm} - \left(\bar{w}^{(gg)}_{12}(k_b,k_a) +\bar{w}^{(gg)}_{12}(k_a,k_b)\right)Z_{1(ab)}[k_{t(ab)};\{u\}]
                                                                              Z_{(ab)2}[k_{t(ab)};\{u\}]u(k_{(ab)})\right]\,,\notag
\end{align}
\begin{align}
\label{eq:DCint}
 {\mathbb K}_{\rm int}^{\rm DC}&[Z[Q;u],u] = \int [d k_a] \int [d k_b] \abar^2(k_{ta}) \Theta(Q-k_{ta}) \Theta(k_{ta}-k_{tb}) \frac{\Delta_{12}(Q)}{\Delta_{12}(k_{ta})}\\
&\hspace{0.3cm}\times\biggl[w^{(0)}_{12}(k_a) \left(w^{(0)}_{1a}(k_b) -\frac{1}{2}
  w^{(0)}_{12}(k_b)\right)  Z_{1b}[k_{ta};\{u\}]
  Z_{ba}[k_{ta};\{u\}]Z_{a2}[k_{ta};\{u\}]u(k_a)u(k_b) \notag\\
&\hspace{0.3cm}+ w^{(0)}_{12}(k_a) \left(w^{(0)}_{a2}(k_b)
  -\frac{1}{2} w^{(0)}_{12}(k_b)\right) Z_{1a}[k_{ta};\{u\}]
  Z_{ab}[k_{ta};\{u\}]Z_{b2}[k_{ta};\{u\}] u(k_a)u(k_b)\notag\\
&\hspace{0.3cm}-w^{(0)}_{12}(k_a) \left(w^{(0)}_{1a}(k_b)  +w^{(0)}_{a2}(k_b)- w^{(0)}_{12}(k_b)\right) Z_{1a}[k_{ta};\{u\}]
  Z_{a2}[k_{ta};\{u\}]u(k_a)\biggr]\notag\,.
\end{align}
Eq.~\eqref{eq:integral_NLL} defines the NLL generating functional for
the non-global evolution. At NLL, at most one gluon at a time inside
the interjet rapidity gap is considered in
Eqs.~\eqref{eq:RRint},~\eqref{eq:DCint}, with configurations with
multiple gluons giving rise to at most NNLL corrections. This
constraint is implicitly enforced in all evolution
equations~\eqref{eq:NLL-evolution-kt-diff} and~\eqref{eq:integral_NLL}
given up to this point.
When taking the four-dimensional limit, one has to supplement
Eq.~\eqref{eq:integral_NLL} with a non-perturbative prescription to
regulate the Landau singularity. We will assume
Eq.~\eqref{eq:initial-cond-4D-alt}, which will be understood in the
following, but alternative models can be adopted.

At LL accuracy, the evolution equation for the generating
functional~\eqref{eq:integral_NLL} is drastically simplified and it
becomes
\begin{align}
\label{eq:integral_Z0}
Z_{12}[Q;\{u\}] = \Delta_{12}(Q) &+  \int [d k_a] \abar(k_{ta}) w^{(0)}_{12}(k_a)
                  \,\frac{\Delta_{12}(Q)}{\Delta_{12}(k_{ta})}\notag\\
  &\times Z_{1a}[k_{ta};\{u\}]
    Z_{a2}[k_{ta};\{u\}] u(k_a) \Theta(Q-k_{ta})\,,
\end{align}
with the LL Sudakov form factor given by
\begin{equation}
\label{eq:sudakov_LL_def}
\ln\Delta_{12}(Q) = - \int [d k] \Theta(Q-k_{t})
\abar(k_{t}) w^{(0)}_{12}(k) \,.
\end{equation}

Before introducing an algorithmic solution of
Eqs.~\eqref{eq:integral_NLL},~\eqref{eq:integral_Z0}, we make some
considerations that will be exploited in their numerical
implementation.
%
We start by considering the term proportional to the independent
emission squared amplitude $w_{12}^{(0)}(k_a) w_{12}^{(0)}(k_b)$ in
Eqs.~\eqref{eq:DCint} and~\eqref{eq:RRint}. According to
Eq.~\eqref{eq:ktij-def} the dipole transverse momentum $k_{tb}$ in
this term is meant to be relative to the $\{12\}$ dipole,
i.e.\ $k_{tb}=k_{tb}^\prime$ by definition for this contribution.
We now express $\bar{w}_{12}^{(gg)}$ in~\eqref{eq:RRint} according to its definition
Eq.~\eqref{eq:double-soft-real}
%
%
and we consider the terms in ${\mathbb K}_{\rm int}^{\rm RR}$
containing $w_{12}^{(0)}(k_a) w_{12}^{(0)}(k_b)$, namely (with
$k_{tb}=k_{tb}^\prime$)
\begin{align}
\label{eq:RRindepint}
 {\mathbb K}_{\rm int}^{\rm RR,~indep.}&[Z[Q;u],u] =  -\int [d k_{a}]\int
                      [d k_{b}] \,\abar^2(k_{t(ab)})\Theta(Q-k_{t(ab)}) {\Theta(k_{ta}-k_{tb})}\frac{\Delta_{12}(Q)}{\Delta_{12}(k_{t(ab)})}\notag\\
&\hspace{-1cm}\times w_{12}^{(0)}(k_a) w_{12}^{(0)}(k_b)\left[ \frac{1}{2}  Z_{1b}[k_{t(ab)};\{u\}]
  Z_{ba}[k_{t(ab)};\{u\}]Z_{a2}[k_{t(ab)};\{u\}]u(k_a)u(k_b)\right. \notag\\
&\left.\hspace{2.05cm}+ \frac{1}{2}  Z_{1a}[k_{t(ab)};\{u\}]
  Z_{ab}[k_{t(ab)};\{u\}]Z_{b2}[k_{t(ab)};\{u\}]u(k_a)u(k_b)\right.\notag\\
&\left. \hspace{2.05cm}\, -  Z_{1(ab)}[k_{t(ab)};\{u\}]
                                                                              Z_{(ab)2}[k_{t(ab)};\{u\}]u(k_{(ab)})\right]\,.
\end{align}
We observe that the difference between the above equation and the
corresponding term proportional to
$w_{12}^{(0)}(k_a) w_{12}^{(0)}(k_b)$ in Eq.~\eqref{eq:DCint} is
logarithmically subleading, contributing at most a NNLL
correction. That is to say that the emission of independent soft
gluons is already correctly iterated by Eq.~\eqref{eq:Uvirtualint} in
the kinematic regime that is relevant to NLL.
This can be understood from simple power counting arguments. Let us
first consider the difference between the double real contributions,
i.e.\ those proportional to the product of two probing functions
$u(k_a) u(k_b)$. Unlike the rest of the double real corrections in
Eqs.~\eqref{eq:RRint},~\eqref{eq:DCint} this term is infrared finite
and it is non-zero only if the two emissions have commensurate
transverse momenta in the $\{12\}$ dipole frame. This condition,
together with the fact that the observable is insensitive to the
region in which $k_a$ and $k_b$ are collinear to the $\{12\}$ dipole
extremities implies that this term yields a relative
${\cal O}(\abar^2)$ correction to the integral equation with no
further logarithmic enhancement, that is NNLL.
Analogous considerations hold for the difference between the collinear
counter-terms in Eqs.~\eqref{eq:DCint},~\eqref{eq:RRindepint}, and
allow us to conclude that all terms proportional to
$w_{12}^{(0)}(k_a) w_{12}^{(0)}(k_b)$ in the r.h.s. of
Eq.~\eqref{eq:integral_NLL} can be neglected for the observables under
consideration.

\begin{figure}[htbp]
 \centering
 \includegraphics[trim={0 8cm 2cm 0}, clip]{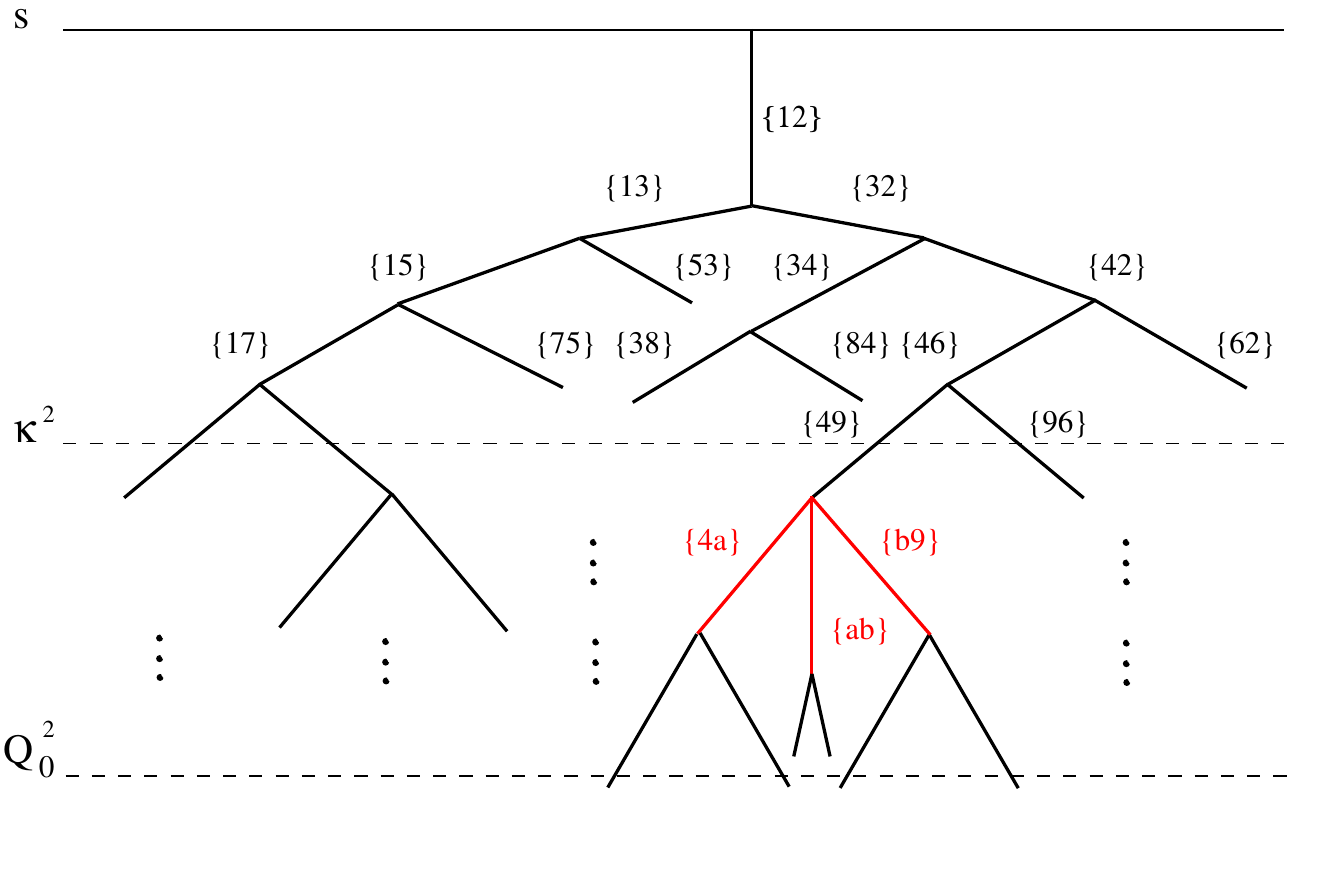}
 \caption{Example of an evolution tree of dipole $\{12\}$. Each edge
   in the graph corresponds to a dipole. The nodes correspond to an
   iteration of $Z^{(0)}[Q;\{u\}]$ except for the (red) node
   describing the splitting of dipole $\{49\}$ which indicates the
   double emission insertion in the r.h.s.\ of the evolution
   equation~\eqref{eq:integral_NLL_Z1} for $Z^{(1)}[Q;\{u\}]$.}
 \label{fig:tree}
\end{figure}

We now observe that the integrated kernel
$ {\mathbb K}_{\rm int}^{\rm RV+VV}[Z[Q;u],u] $ in
Eq.~\eqref{eq:integral_NLL} has the same functional form as the LL
equation~\eqref{eq:integral_Z0}, and it can be solved with the same
algorithm (Algorithm~\ref{algo:LLevolution} in
Sec.~\ref{sec:PT-solution}).
It is therefore appropriate to split $Z_{12}[Q;\{u\}]$ into the sum of
two contributions as~\footnote{The strategy that follows is inspired
  to what has been already applied to the NNLL calculation of global
  QCD
  observables~\cite{Banfi:2012jm,Banfi:2014sua,Banfi:2016zlc,Banfi:2018mcq,Monni:2019yyr},
  and specifically to the insertion of a \textit{correlated} pair of
  soft partons.}
\begin{equation}
\label{eq:pert}
Z_{12}[Q;\{u\}] = Z_{12}^{(0)}[Q;\{u\}] + Z_{12}^{(1)}[Q;\{u\}]\,,
\end{equation}
where $Z_{12}^{(0)}[Q;\{u\}]$ satisfies the integral equation
\begin{align}
\label{eq:integral_NLL_Z0}
Z_{12}^{(0)}&[Q;\{u\}] = \Delta_{12}(Q) +  \int [d k_a] \abar(k_{ta}) w^{(0)}_{12}(k_a)
                  \,\bigg(1 +\abar(k_{ta})\,\bar{K}^{(1)}\bigg)\frac{\Delta_{12}(Q)}{\Delta_{12}(k_{ta})}\notag\\
  &\hspace{0.5cm}\times Z_{1a}^{(0)}[k_{ta};\{u\}]
    Z_{a2}^{(0)}[k_{ta};\{u\}] u(k_a) \Theta(Q-k_{ta})\,,
\end{align}
with $\Delta_{12}$ defined in Eq.~\eqref{eq:sudakov_NLL_def}.
We can now treat $Z_{12}^{(1)}[Q;\{u\}]$ as a perturbation, observing
that all contributions to $\Sigma(v)$ that are quadratic in
$Z_{12}^{(1)}[Q;\{u\}]$ (as well as those proportional to
$Z_{12}^{(1)}[Q;\{u\}] \bar{K}^{(1)}$) only correspond to NNLL corrections.
We can therefore linearise Eq.~\eqref{eq:integral_NLL} in
$Z_{12}^{(1)}[Q;\{u\}]$ by inserting Eq.~\eqref{eq:pert} into
Eq.~\eqref{eq:integral_NLL} and neglecting the aforementioned
quadratic corrections:
%
\begin{align}
\label{eq:integral_NLL_Z1}
Z^{(1)}_{12}&[Q;\{u\}] \simeq \int [d k_a] \abar(k_{ta}) w^{(0)}_{12}(k_a)
                  \,\frac{\Delta_{12}(Q)}{\Delta_{12}(k_{ta})}\\
  &\hspace{0.5cm}\times \left(Z^{(0)}_{1a}[k_{ta};\{u\}]
    Z^{(1)}_{a2}[k_{ta};\{u\}] + Z^{(1)}_{1a}[k_{ta};\{u\}]
    Z^{(0)}_{a2}[k_{ta};\{u\}] \right) u(k_a) \Theta(Q-k_{ta})\notag\\
  &+ \int [d k_{a}]\int
                      [d k_{b}] \,\abar^2(k_{t(ab)})\Theta(Q-k_{t(ab)}) {\Theta(k_{ta}-k_{tb}^\prime)}\frac{\Delta_{12}(Q)}{\Delta_{12}(k_{t(ab)})}\notag\\
&\hspace{0.5cm}\times\left[ \tilde{w}^{(0)}_{12}(k_b,k_a)  Z^{(0)}_{1b}[k_{t(ab)};\{u\}]
  Z^{(0)}_{ba}[k_{t(ab)};\{u\}]Z^{(0)}_{a2}[k_{t(ab)};\{u\}]u(k_a)u(k_b)\right. \notag\\
&\left.\hspace{0.5cm}+\, \tilde{w}^{(0)}_{12}(k_a,k_b) Z^{(0)}_{1a}[k_{t(ab)};\{u\}]
  Z^{(0)}_{ab}[k_{t(ab)};\{u\}]Z^{(0)}_{b2}[k_{t(ab)};\{u\}]u(k_a)u(k_b)\right.\notag\\
&\left. \hspace{0.5cm} -\, \left(\tilde{w}^{(0)}_{12}(k_b,k_a) +\tilde{w}^{(0)}_{12}(k_a,k_b)\right)Z^{(0)}_{1(ab)}[k_{t(ab)};\{u\}]
                                                                              Z^{(0)}_{(ab)2}[k_{t(ab)};\{u\}]u(k_{(ab)})\right]\notag\\
&-\int [d k_a] \int [d k_b] \abar^2(k_{ta}) \Theta(Q-k_{ta}) \Theta(k_{ta}-k_{tb}) \frac{\Delta_{12}(Q)}{\Delta_{12}(k_{ta})}\notag\\
&\hspace{0.5cm}\times\biggl[w^{(0)}_{12}(k_a) w^{(0)}_{1a}(k_b)  Z^{(0)}_{1b}[k_{ta};\{u\}]
  Z^{(0)}_{ba}[k_{ta};\{u\}]Z^{(0)}_{a2}[k_{ta};\{u\}]u(k_a)u(k_b) \notag\\
&\hspace{0.5cm}+ \,w^{(0)}_{12}(k_a) w^{(0)}_{a2}(k_b) Z^{(0)}_{1a}[k_{ta};\{u\}]
  Z^{(0)}_{ab}[k_{ta};\{u\}]Z^{(0)}_{b2}[k_{ta};\{u\}] u(k_a)u(k_b)\notag\\
&\hspace{0.5cm}-\,w^{(0)}_{12}(k_a) \left(w^{(0)}_{1a}(k_b)  +w^{(0)}_{a2}(k_b)\right) Z^{(0)}_{1a}[k_{ta};\{u\}]
  Z^{(0)}_{a2}[k_{ta};\{u\}]u(k_a)\biggr]\notag\,,
\end{align}
where $\simeq$ indicates that we neglect corrections of order NNLL and
higher.
Eq.~\eqref{eq:integral_NLL_Z1} has a simple physical
interpretation. The two double integrals in the r.h.s.\ of
Eq.~\eqref{eq:integral_NLL_Z1} correspond to an \textit{insertion} of
a pair of soft gluons $k_a$, $k_b$ with commensurate energies (or
transverse momenta in the $\{12\}$ frame), but strongly ordered
w.r.t.\ the rest of the soft radiation in the evolution.
The emission of the unordered pair will give rise to three colour
dipoles (two for the corresponding collinear counter-terms), which will
subsequently evolve according to the evolution
equation~\eqref{eq:integral_NLL_Z0}.
The first term in the r.h.s.\ of Eq.~\eqref{eq:integral_NLL_Z1} states
that such insertion can occur at \textit{any} stage of the evolution,
i.e.\ the pair $k_a$, $k_b$ does not necessarily correspond to the
first two soft gluons emitted off the initial $q\bar{q}$ dipole. This
structure is depicted in the evolution tree of Fig.~\ref{fig:tree},
where the the edges of the graph correspond to colour dipoles. In this
example, each node corresponds to the splitting of a dipole into two
adjacent dipoles according to Eq.~\eqref{eq:integral_NLL_Z0}, with the
exception of the (red) splitting of the dipole $\{49\}$ which depicts
the double emission insertion in the r.h.s.\ of
Eq.~\eqref{eq:integral_NLL_Z1}. This $Z^{(1)}[Q;\{u\}]$ insertion can
occur at any branching, and one needs therefore to sum over all
possible configurations.
This procedure is a perturbative solution of
Eq.~\eqref{eq:integral_NLL}, where we neglect corrections that are
subleading (i.e.\ NNLL) in the strict logarithmic counting. One could
also envision an algorithmic solution of the full integral equation,
in which roughly speaking the insertion of $Z^{(1)}[Q;\{u\}]$ is
iterated an arbitrary number of times. The corresponding algorithm
would differ from the solution presented here by NNLL corrections, and
it would be directly relevant for the inclusion of NLL non-global
corrections in the frame of a full-fledged dipole shower.  For this
reason, its formulation is left for future investigations.

The above equations can be solved recursively, for which Monte Carlo
techniques offer a natural theoretical tool.
In the next section we will introduce the necessary algorithms to
perform the NLL resummation using the technology of dipole showers
ordered in the dipole transverse momentum.
Algorithms of this type are very common in the parton shower
literature (see
e.g.\ Refs.~\cite{Gustafson:1987rq,Lonnblad:1992tz,Sjostrand:2004ef,Giele:2007di,Hoche:2015sya,Dasgupta:2020fwr,Forshaw:2020wrq}),
and at present they usually achieve LL accuracy for the observables
considered in this article. Notice, however, that in the present paper
these Monte Carlo techniques are simply used to solve numerically the
evolution equations and not to build an actual generator of physical
events.

\section{Perturbative solution of the evolution equations}
\label{sec:PT-solution}
In this section we discuss how Eq.~\eqref{eq:master_genfun} can be
calculated using Monte Carlo methods.
We will start with Sec.~\ref{sec:ll-evolution} by reporting the
dipole-shower algorithm to solve the LL counter-part of
Eq.~\eqref{eq:integral_NLL}, originally derived in
Ref.~\cite{Dasgupta:2001sh}, then we move on to discuss the NLL
corrections to the cumulative cross section~\eqref{eq:master} (or
equivalently~\eqref{eq:master_genfun}), in configurations with three
(Sec.~\ref{sec:3jet}) and two (Sec.~\ref{sec:nll-2jet}) hard partons
in the final state.

In the rest of this article, we also restore the full dependence on
$N_c$ both in the QCD beta function used in the evolution of the
running coupling~\eqref{eq:RGEplanar} and in the cusp anomalous
dimension~\eqref{eq:cusp}. This by no means implies full control over
finite-$N_c$ corrections, as additional subleading-$N_c$ terms are
neglected in the evolution equations. We therefore replace
Eq.~\eqref{eq:RGEplanar} with the full RGE (notice that an overall
factor $N_c/\pi$ is present in Eq.~\eqref{eq:RGEplanar})
\begin{equation}
\label{eq:RGEfull}
\frac{d\alpha_s(\mu)}{d\ln\mu^2} = -\beta(\alpha_s) = -\alpha_s
\left(\beta_0 \alpha_s + \beta_1 \alpha_s^2 + \dots\right)\,,
\end{equation}
with 
\begin{equation}
\beta_0 = \frac{11 C_A-2
n_F}{12\pi}\,,\quad \beta_1 = \frac{17 C_A^2 - 5 C_A n_F - 3 C_F n_F}{24 \pi^2}\,.
\end{equation}
Moreover we replace in Eq.~\eqref{eq:integral_NLL}
\begin{equation}
\abar(k_{ta})\bar{K}^{(1)} \to \frac{\alpha_s(k_{ta})}{2\pi}\,K^{(1)}\,,\quad K^{(1)} =
\left(\frac{67}{18}-\frac{\pi^2}{6}\right) C_A- \frac{5}{9}n_F \,.
\end{equation}
Similarly, we will retain the full colour quadratic $SU(N_c)$ Casimir
operators in the calculation of the hard coefficients ${\cal H}_{2}$
and ${\cal H}_3$ in Sec.~\ref{sec:3jet} below.

\subsection{LL evolution algorithm}
\label{sec:ll-evolution}
At LL, Eq.~\eqref{eq:integral_Z0} can be solved with the dipole shower
algorithm derived in the pioneering article by Dasgupta and
Salam~\cite{Dasgupta:2001sh}, that we adapt below to the problem
considered here and to the case of dipole-transverse-momentum
ordering.
For later use, we define the \textit{evolution time}
\begin{equation}
\label{eq:time}
t \coloneqq \int_{\frac{k_t}{\sqrt{s}}}^1\frac{d x}{x} \abar(x \sqrt{s}) =-
\frac{N_c}{2\pi\beta_0}\ln \left(1-2 \lambda\right)\,,\qquad \lambda=\beta_0\alpha_s(\sqrt{s})\ln\frac{\sqrt{s}}{k_t}\,.
\end{equation}
The solution to Eq.~\eqref{eq:integral_Z0} is then obtained with
Algorithm~\ref{algo:LLevolution}, which is iterated until the desired
statistical precision is reached.
In the following, we define the infrared scale $Q_0$ as the
singularity of Eq.~\eqref{eq:time}, namely such that
\begin{equation}
\label{eq:Q0def}
2\,\beta_0\alpha_s(\sqrt{s})\ln\frac{\sqrt{s}}{Q_0} = 1\,.
\end{equation}
In practice, this scale is extremely low and therefore the evolution
is rarely stopped because the scale $Q_0$ is reached before any
emission is radiated into the interjet rapidity gap.

An important aspect to stress about Algorithm~\ref{algo:LLevolution},
is that at every step emissions are generated in the emitting dipole
rest frame, but the observable is calculated in the event frame.
We therefore need to apply a simple Lorentz transformation at every
evolution step to transform the generated emission into the event
frame.
In doing this, we only need to keep track of the direction of the
generated momenta, while the information regarding the \textit{dipole}
transverse momentum (i.e. the normalisation of the momenta) is encoded
in the evolution time $t$.
Therefore, we divide all momenta by their energy in the event frame,
and keep track of the normalisation separately.
As in Ref.~\cite{Dasgupta:2001sh}, all Lorentz transformation
discussed in all algorithms presented in this section have to be
performed with \textit{normalised} momenta.
This solves the problem of handling numerically Lorentz
transformations involving very soft momenta. Also, it allows us to
perform the evolution without any momentum conservation at any
evolution step, thus eliminating exactly all subleading power
(i.e. non-logarithmic) corrections.
\begin{algorithm}
   Set $i=0$ and the evolution time $t_0=0$\;
   Start with one initial dipole made by the Born fermionic line\;
   \While{{\rm true}}
   {
      Compute $\Delta\eta_{\rm tot}=\sum_{\ell=1}^N\Delta\eta_{\ell}$, the sum of the available rapidity
      ranges within each of the $N$ dipoles in the event so far\;
     Increase $i$ by 1\;
     Generate a random number $r \in [0,1]$ \& increase $t$ by an amount $\Delta t = t_i-t_{i-1}$ generated by solving
     $$\frac{\Delta_{12}(k_{t,i-1})}{\Delta_{12}(k_{t,i})} = e^{-\Delta\eta_{\rm
    tot}\Delta t} = r\,;$$
Generate the \textit{dipole} transverse momentum $k_{t,i}$ of the next
emission $k_i$ by solving Eq.~\eqref{eq:time} with the new
$t_i$. Generate its azimuth uniformly in $[0,2 \pi]$ and its rapidity
such that the emission is at a safety rapidity distance $\delta$ from
the dipole extremities in the \textit{event} frame\;
     Choose the emitting dipole ${\cal D}_\ell$ with
     probability $\Delta\eta_\ell/\Delta\eta_{\rm tot}$\;
     \If{$k_{t,i} < Q_0$} {break\;}
     Split the dipole ${\cal D}_\ell$ into two adjacent dipoles\;
     \If{$\Theta_{\rm in}(k_i) = 1$} {Calculate observable and add the event to the histogram
       \& break\;}
   }
\caption{LL evolution algorithm}
\label{algo:LLevolution}
\end{algorithm}
We now discuss the calculation of the observable in a given event. The
evolution in Algorithm~\ref{algo:LLevolution} stops as soon as one
gluon is emitted inside the interjet rapidity gap. We then calculate
the transverse energy $E_t$ of this gluon w.r.t. the thrust axis of
the event, and add the event to the histogram.
We notice that in Ref.~\cite{Dasgupta:2002bw}, the transverse energy
is instead defined as the value of the ordering variable. This
definition is correct at LL, but the exact relation between the
ordering variable and the actual observable leads to a genuine NLL
effect.
In this article we include these effects already in the LL prediction,
and therefore we work with the physical observable everywhere in our
calculation.

\subsection{NLL evolution algorithm: ${\cal H}_3\otimes S_3(v)$ contribution}
\label{sec:3jet}
We now move on with the NLL corrections. We start from the
${\cal H}_3\otimes S_3(v)$ contribution to the cumulative cross
section~\eqref{eq:master}, which describes the production of three
hard partons inside the jets (i.e.\ outside the interjet rapidity gap).
The three partons can be viewed as two independent colour dipoles in a
large-$N_c$ picture (in the case of $e^+e^-$ collisions), labelled as
$\{13\}$ and $\{23\}$ in the second term of the r.h.s.\ of
Eq.~\eqref{eq:master_genfun}. These subsequently emit soft
radiation independently of each other into the measured interjet
region.
The soft evolution of each of the above two dipoles, encoded in $S_3$,
is carried out at LL order using Algorithm~\ref{algo:LLevolution}, and
it is therefore straightforward.
However, some care must be taken in the calculation of the hard factor
${\cal H}_3$ corresponding to the three hard-parton
contribution. Specifically, in Ref.~\cite{Banfi:2021owj} as well as in
Sec.~\ref{sec:recall} we have stated that the hard factors are
individually IRC finite, while the integral over the three-parton
final state clearly has divergences associated to singular kinematic
configurations.
At NLL, these divergence are meant to be cancelled by corresponding
divergent contributions in the virtual corrections entering
${\cal H}_2$, and such a cancellation has to be enforced by means of a
subtraction procedure. This also implies that the precise definition
of the hard matching coefficients ${\cal H}_2$ and ${\cal H}_3$
depends on the scheme adopted to subtract their IRC divergences and
only their combination has a physical meaning.
In Section 5 of Ref.~\cite{Banfi:2021owj} the calculation was carried
out analytically. However, in this article we would like to take a
different approach and set up a numerical calculation using a local
subtraction method that can be easily applied to the case of more
complicated processes. As a consequence, the individual definition of
${\cal H}_2$ and ${\cal H}_3$ computed here will differ from those of
Ref.~\cite{Banfi:2021owj} while their physical sum will be identical.

We start by labelling with $p_1$, $p_2$, $p_3$ the quark, antiquark
and gluon respectively. With the usual $x_1,x_2,x_3$ variables
\begin{equation}
  \label{eq:xi}
  x_i = \frac{2 (p_i \cdot q)}{s}\,,\quad i=1,2,3\,,\qquad x_1+x_2+x_3=2\,,
\end{equation}
where $q$ is the four momentum of the virtual photon $q^\mu =
(\sqrt{s},\vec{0})$.
To obtain ${\cal H}_3\otimes S_3(v)$, we start by evaluating the
integral over the $q\bar{q}g$ phase space. We choose the reference
frame so that the $z$ axis is along the direction of the quark
$\vec{p}_1$, and we explicitly parametrise the phase space of the
remaining two partons in terms of the energy fraction of the gluon
$x_3 = 2E_g/\sqrt{s}$ and the cosine of the angle between the gluon
and the quark $y=\cos\theta_{qg}$. The real contribution to
${\cal H}_3\otimes S_3(v)$ is
\begin{align}
  \label{eq:Sigmav-3jet}
\Sigma^{{\rm real}}(v) &= 
  2C_F\frac{\alpha_s}{2\pi}\left(\frac{\mu^2}{s}\right)^{\epsilon}\frac{e^{\gamma_E\epsilon}}{\Gamma(1-\epsilon)}\int_0^1d
                                      x_3 \frac{x_3^{-1-2
                                      \epsilon}}{(1-x_3)^{2
                                      \epsilon}}\int_{-1}^1 d y\,  \frac{(1-y)^{-1-\epsilon} (1+y)^{-1-\epsilon}}{ (2-x_3
   (1-y))^{2 -2\epsilon} }\notag\\
&\times\left(8-\epsilon x_3^2 (2-x_3 (1-y))^2-(2-x_3) x_3 \left((x_3-2) x_3
                                     (1-y)^2-4
                                     y+8\right)\right)\notag\\
  &\times \Theta_{\rm out}(p_1) \Theta_{\rm out}(p_2) \Theta_{\rm out}(p_3) \, S_{3}(v)\,,
\end{align}
where the coupling has been renormalised in the $\overline{\rm MS}$
scheme.
The phase space constraint $\Theta_{\rm out}(p_1) \Theta_{\rm
  out}(p_2) \Theta_{\rm out}(p_3)$ is non trivial, and imposes that
none of the hard particles ends up inside the interjet rapidity gap,
as per definition of ${\cal H}_3$. We stress that the direction of the
thrust axis, that is used to define the position of the interjet
rapidity gap, is now aligned with the hardest parton.
For a given value of $x_3$ and $y$ the event is then dressed by a
shower of soft gluons encoded in $S_3$, so that the integration over
the remaining phase space of the three-parton system (specifically
$y$) involves also the soft factor $S_3$.
Eq.~\eqref{eq:Sigmav-3jet} produces double and single poles of soft
and collinear origin, and we wish to perform a local subtraction of
these divergences so that the above integral is computed numerically.
We consider a simple subtraction scheme in which the local counter-term
is defined by the full real integrand albeit with unresolved
kinematics. That is, we build the counter-term by replacing the phase
space constraint in Eq.~\eqref{eq:Sigmav-3jet} with
\begin{equation}
\Theta_{\rm out}(p_1) \Theta_{\rm out}(p_2) \Theta_{\rm
  out}(p_3) \to \Theta_{\rm
  out}^{\rm soft}(p_3)\,.
\end{equation}
Here $\Theta_{\rm out}^{\rm soft}(p_3)$ indicates that when the gluon
$p_3$ is unresolved (i.e.\ either soft or collinear to either quark
leg) the thrust axis is aligned along the $z$-axis and therefore
$\Theta_{\rm out}(p_1) = \Theta_{\rm out}(p_2) =1$ by
construction. The addition of such a counter-term modifies
Eq.~\eqref{eq:Sigmav-3jet} as follows
\begin{align}
  \label{eq:Sigmav-3jet-sub}
\Sigma^{(3), {\rm sub}}(v) &= 
  2C_F\frac{\alpha_s}{2\pi}\left(\frac{\mu^2}{s}\right)^{\epsilon}\frac{e^{\gamma_E\epsilon}}{\Gamma(1-\epsilon)}\int_0^1d
                                      x_3 \frac{x_3^{-1-2
                                      \epsilon}}{(1-x_3)^{2
                                      \epsilon}}\int_{-1}^1 d y\,  \frac{(1-y)^{-1-\epsilon} (1+y)^{-1-\epsilon}}{ (2-x_3
   (1-y))^{2 -2\epsilon} }\notag\\
&\times\left(8-\epsilon x_3^2 (2-x_3 (1-y))^2-(2-x_3) x_3 \left((x_3-2) x_3
                                     (1-y)^2-4
                                     y+8\right)\right)\notag\\
  &\times \left[\Theta_{\rm out}(p_1) \Theta_{\rm out}(p_2)
    \Theta_{\rm out}(p_3) \, S_{3}(v) - \Theta_{\rm
    out}^{\rm soft} (p_3) \, S_{2}(v) \right]\,,
\end{align}
where $S_2$ in the last term indicates that the soft factor now does
not see the unresolved gluon $p_3$ and therefore it degenerates into
the two-leg factor $S_2$. 
We point out that this procedure is a simple adaptation of the
\textit{projection-to-Born} subtraction method~\cite{Cacciari:2015jma}
to an all-order calculation, where the projection acts on the full
real phase space including the soft factor $S_3$.
Eq.~\eqref{eq:Sigmav-3jet-sub} is free of collinear singularities,
however it still contains a soft singularity due to the fact that
$S_{3}(v)$ depends on the direction of $p_3$ regardless of how soft
the latter is. We then introduce a technical cutoff on the
transverse momentum of the gluon $p_3$ w.r.t. the $\{12\}$ dipole
($p_{t,3}^{\{12\}} > Q_0$), and we set $\epsilon \to 0$ and evaluate
the integral with Algorithm~\ref{algo:NLLevolution3j} (again
iterated until the desired statistical precision is reached).  
\begin{algorithm}
  Generate $x_3$ and $y$ and parametrise the kinematics of the
  $q\bar{q}g$ system in terms of these variables\;
  \If{$p_{t,3}^{\{12\}} < Q_0$} {break and generate a new event\;}
  Set the weight $w$ to the integrand in the first two lines of
  Eq.~\eqref{eq:Sigmav-3jet-sub} with $\epsilon=0$\;
  Create an \textit{event}:\\
     \While{{\rm true}}
     {
       Set the thrust axis along the direction of the hardest parton\;       
       \If{$\Theta_{\rm out}(p_1) \Theta_{\rm out}(p_2)
        \Theta_{\rm out}(p_3) = 0$} {break\;}
       Consider the two large-$N_c$ dipoles $\{13\}$ and $\{23\}$
       and apply the LL evolution algorithm~\ref{algo:LLevolution} to compute
       the soft factor $S_3$ at leading colour\;
       break\;
     }

     Create a \textit{counter-event}:\\
     Set $w \mathrel{*}= -1$\;
     \While{{\rm true}}
     {
       Set the thrust axis along the direction of the quark ($z$ direction)\;
       \If{$\Theta_{\rm out}^{\rm soft} (p_3) = 0$} {break\;}
       Consider the dipole $\{12\}$ and apply the LL evolution algorithm~\ref{algo:LLevolution} to compute
       the soft factor $S_2$ while filling the same histogram as for the event\;
       break\;
     }
\caption{NLL evolution algorithm for ${\cal H}_3\otimes S_3(v)$}
\label{algo:NLLevolution3j}
\end{algorithm}
%
The computation of Eq.~\eqref{eq:Sigmav-3jet-sub} does not directly
return the contribution ${\cal H}_3\otimes S_3(v)$. This is obtained
by subtracting the double counting with the term
${\cal H}_2\otimes S_2(v)$, where the first gluon $p_3$ is now
generated according to the LL evolution kernel. This requires
subtracting from Eq.~\eqref{eq:Sigmav-3jet-sub} the
term~\cite{Banfi:2021owj} (we set $N_c\rightarrow 2 C_F$ for this
first emission)
\begin{align}
  \label{eq:Sigmav-3jet-soft-sub}
\Sigma^{(3) , {\rm sub}}_{\rm soft}(v) &= 4C_F\frac{\alpha_s}{2\pi}\mu^{2\epsilon}\frac{e^{\gamma_E\epsilon}}{\Gamma(1-\epsilon)}\int_{0}^{\sqrt{s}}\frac{d
    k_t}{k_t^{1+2\epsilon}}\int_{\ln (k_t/\sqrt{s})}^{\ln
  (\sqrt{s}/k_t)}\,d \eta\,\Theta^{\rm soft}_{\rm out}(k)
\left[S^{\rm soft}_3(v) -S_2(v)\right]\,,
\end{align}
where $S^{\rm soft}_3$ indicates that the emission of the soft gluon
$k$ does not cause any recoil in the $q\bar{q}g$ event kinematics, and
therefore the thrust axis is always aligned with the $z$-axis.
In Eq.~\eqref{eq:Sigmav-3jet-soft-sub} we have adopted the same
subtraction method used in Eq.~\eqref{eq:Sigmav-3jet-sub}, thereby
subtracting the local counter-term evaluated in the unresolved
(i.e.\ two-leg) kinematics. 
Eq.~\eqref{eq:Sigmav-3jet-soft-sub} also contains the soft singularity
present in Eq.~\eqref{eq:Sigmav-3jet-sub}, and therefore we need to
apply the same technical cutoff $k_t > Q_0$ here. It is now crucial to
notice that the difference between the two equations is instead finite
in the limit $Q_0\to 0$, and the regulator can be pushed to negligibly
small values in the combination of the two.
  Eq.~\eqref{eq:Sigmav-3jet-soft-sub} can be evaluated with a slightly
modified version of Algorithm~\ref{algo:NLLevolution3j}, given in
Algorithm~\ref{algo:NLLevolution3jsoft}.
\begin{algorithm}
  Generate $\vec{k}_t$ and $\eta$ of the gluon $k$ and the back-to-back
  $q\bar{q}$ pair along the $z$ axis\;
  \If{$k_t < Q_0$} {break and generate a new event\;}
  Set the weight $w$ to the integrand in Eq.~\eqref{eq:Sigmav-3jet-soft-sub} with
  $\epsilon=0$ (w/o $\Theta$ and $S_n$ factors)\;
  Set the thrust axis along the direction of the quark ($z$ direction)\;
  Create an \textit{event}:\\
     \While{{\rm true}}
     {
       \If{$\Theta_{\rm out}^{\rm soft} (k) = 0$} {break\;}
       Consider the two large-$N_c$ dipoles $\{13\}$ and $\{23\}$
       and apply the LL evolution algorithm~\ref{algo:LLevolution} to compute
       the soft factor $S^{\rm soft}_3$ at leading-colour\;
       break\;
     }

     Create a \textit{counter-event}:\\
     Set $w \mathrel{*}= -1$\;
     \While{{\rm true}}
     {
       \If{$\Theta_{\rm out}^{\rm soft} (k) = 0$} {break\;}
       Consider the dipole $\{12\}$ and apply the LL evolution algorithm~\ref{algo:LLevolution} to compute
       the soft factor $S_2$ while filling the same histogram as for the event\;
       break\;
     }
\caption{NLL evolution algorithm for the soft contribution to ${\cal H}_3\otimes S_3(v)$}
\label{algo:NLLevolution3jsoft}
\end{algorithm}
Finally we obtain ${\cal H}_3\otimes S_3(v)$ as the
\textit{difference} between the two contributions
\begin{equation}
\label{eq:H3S3_final}
{\cal H}_3\otimes S_3(v) = \Sigma^{(3) , {\rm sub}}(v) - \Sigma^{(3) , {\rm sub}}_{\rm soft}(v) \,.
\end{equation}
The above procedure used to define ${\cal H}_3$ implicitly also
defines uniquely the two-parton hard coefficient ${\cal H}_2$.
This will be given by the one loop correction to the quark form factor
minus the integrals of the local counter-terms appearing in
Eq.~\eqref{eq:H3S3_final}, minus the virtual correction to the
evolution kernel that is subtracted to avoid the double counting with
$S_2$ (see Section 5 of Ref.~\cite{Banfi:2021owj}). The latter
contribution to ${\cal H}_2\otimes S_2(v)$ reads~\cite{Banfi:2021owj}
\begin{align}
\Sigma^{(2),{\rm virt.}}_{\rm soft}(v) &=   4C_F\frac{\alpha_s}{2\pi}\mu^{2\epsilon}\frac{e^{\gamma_E\epsilon}}{\Gamma(1-\epsilon)}\int_{0}^{\sqrt{s}}\frac{d
    k_t}{k_t^{1+2\epsilon}}\int_{\ln (k_t/\sqrt{s})}^{\ln (\sqrt{s}/k_t)}d \eta \,S_2(v)\,.
\end{align}
This yields
\begin{align}
\label{eq:H2}
{\cal H}_2 &=
             \delta^{(2)}(\Omega_1-\Omega_q)\delta^{(2)}(\Omega_2-\Omega_{\bar
             q}) \left(1 \!+\! \frac{\alpha_s}{2\pi}{\cal H}_{2}^{(1)}
             + {\cal O}(\alpha_s^2)\right)\,,
\end{align}
where
\begin{align}
\label{eq:H2_1}
{\cal H}&_2^{(1)} =\frac{C_F}{2
   \left(1-c^2\right)^2}\bigg(4
   \left(1-c^2\right)^2 \left(\text{Li}_2\left(\frac{1+c}{2}\right)\!-\! \text{Li}_2\left(\frac{1-c}{2}\right)\right)\notag\\&-2 \left(1-c^2\right)^2 \log
   ^2(1+c)+16 c \left(3+c^2\right) \ln(2) -(1-c^2) (c (16+3 c)-3)\notag\\&+2 \ln(1-c) \left(-2 \left(1+c^4\right) \log
   (2)-4 c \left(3+c^2\right)+\left(1-c^2\right)^2 \ln(1-c)\right) \\&+\left(4
   \left(1+c^4\right) \ln(2)-8 c \left(3+c^2\right)\right) \ln(1+c)-4 \left(-3 c^4+2
   c^2 (9+2 \ln(2))+1\right) \tanh ^{-1}(c) \bigg)\,.\notag
\end{align}
Here $c$ is the cosine of the jet opening angle, defined in
Eq.~\eqref{eq:c_def}.
This coefficient will be used in the next section for the calculation
of the ${\cal H}_2\otimes S_2(v)$ contribution to the NLL cumulative
cross section.

\subsection{NLL evolution algorithm: ${\cal H}_2\otimes S_2(v)$ contribution}
\label{sec:nll-2jet}

We now address the numerical solution to the evolution
equations~\eqref{eq:integral_NLL_Z0}, \eqref{eq:integral_NLL_Z1}. We
start by considering the contribution $Z^{(0)}_{12}[Q;\{u\}]$, defined
by the evolution equation~\eqref{eq:integral_NLL_Z0}. The solution to
Eq.~\eqref{eq:integral_NLL_Z0} can be obtained with proper
modifications of Algorithm~\ref{algo:LLevolution}. 
However, while Algorithm~\ref{algo:LLevolution} can be used to
calculate $Z^{(0)}_{12}[Q;\{u\}]$, defined by
Eq.~\eqref{eq:integral_NLL_Z0}, in order to include the contribution
from $Z^{(1)}_{12}[Q;\{u\}]$ given in Eq.~\eqref{eq:integral_NLL_Z1}
we cannot simply run Algorithm~\ref{algo:LLevolution} down to the
infrared scale $Q_0$.  
Instead, we first introduce a \textit{truncated} version of
Algorithm~\ref{algo:LLevolution}, given in
Algorithm~\ref{algo:NLLevolution}.

\begin{algorithm}[h]
  Set the thrust axis along the $z$ axis\;
  Generate the truncation scale $\kappa \in [Q_0,\sqrt{s}]$
  uniformly according to eq.~\eqref{eq:kappa}\;
  Set the weight $w={\cal H}_2$ as defined in Eq.~\eqref{eq:H2}\;
  Apply Algorithm~\ref{algo:LLevolution} with evolution
  time~\eqref{eq:time_NLL} truncated at $Q_0=\kappa$\;
\caption{Truncated NLL algorithm for the first evolution branch}
\label{algo:NLLevolution}
\end{algorithm}
%
Firstly, we replace the evolution time~\eqref{eq:time} by its NLL
counterpart
\begin{align}
\label{eq:time_NLL}
t \coloneqq & \int_{\frac{k_t}{\sqrt{s}}}^1\frac{d x}{x} \abar(x \sqrt{s}) \bigg(1 +\frac{\alpha_s(x \sqrt{s})}{2\pi}\,K^{(1)}\bigg)=-
    \frac{N_c}{2\pi\beta_0}\ln \left(1-2\lambda\right)\notag\\
  & +
    \abar(\sqrt{s})\left[\frac{\lambda}{1-2\lambda}\left(\frac{K^{(1)}}{2\pi\beta_0}-\frac{\beta_1}{\beta_0^2}\right)-\frac{\ln(1-2\lambda)}{1-2\lambda}\frac{\beta_1}{2\beta_0^2}\right]
    + {\cal O}({\rm NNLL})\,.
\end{align}
The above equation uniquely defines the dipole transverse momentum
given the evolution time for a given emission.
We now introduce the scale $\kappa$ at which we truncate a first
branch of the evolution carried out according to the
equation~\eqref{eq:integral_NLL_Z0}.
The rationale is to use the simple partition of unity in the evolution
sequence
\begin{equation}
\label{eq:kappa}
1 = \int_{Q_0}^{\sqrt{s}}\frac{d\kappa}{\sqrt{s}-Q_0}\,,
\end{equation}
and use the fact that the evolution~\eqref{eq:integral_NLL_Z0} between
$\sqrt{s}$ and $Q_0$ is identical to the combination of the evolution
between $\sqrt{s}$ and $\kappa$, and the evolution between $\kappa$
and $Q_0$.
For each value of the scale $\kappa$ we will insert either one or two
emissions, to perform a calculation of
Eqs.~\eqref{eq:integral_NLL_Z0},~\eqref{eq:integral_NLL_Z1}. These are
identified by the three contributions defined below.
According to Eq.~\eqref{eq:kappa} this scale is sampled uniformly and
ensures that each insertion can occur at any possible scale along the
evolution tree.
%

\renewcommand{\thealgocf}{\arabic{algocf}.\arabic{subalgo}}
\addtocounter{algocf}{-1}
\addtocounter{subalgo}{1}
\begin{algorithm}[h]
Generate emission $k_a$ according to Algorithm~\ref{algo:ka} and split
the emitting dipole\;
Apply Algorithm~\ref{algo:LLevolution} with starting scale
    $k_{ta}$ and evolution time~\eqref{eq:time_NLL}\;
\caption{Insertion of $Z^{(0)}_{12}[Q;\{u\}]$ starting from the scale $\kappa$}
\label{algo:NLLevolution_Z0}
\end{algorithm}
\addtocounter{algocf}{-1}
\addtocounter{subalgo}{1}
To carry on with the calculation of $Z^{(0)}_{12}[Q;\{u\}]$ starting
from Algorithm~\ref{algo:NLLevolution}, we simply generate an emission
$k_a$ according to Eq.~\eqref{eq:integral_NLL_Z0} and then continue
with the evolution from the scale of the emission all the way down to
$Q_0$. This procedure is described in
Algorithm~\ref{algo:NLLevolution_Z0}.

\begin{algorithm}[htp]
  Select an emitting dipole $\{p_{\ell_1} p_{\ell_2}\}$ for $k_a$ as in Algorithm~\ref{algo:LLevolution}\;
  Generate $k_{ta}$ as in Algorithm~\ref{algo:LLevolution}, starting from
  the scale $\kappa$\;
\caption{Generation of emission $k_a$ starting from the scale $\kappa$}
\label{algo:ka}
\end{algorithm}
%
%
\addtocounter{algocf}{-1}
\addtocounter{subalgo}{1}
Let us now move to Eq.~\eqref{eq:integral_NLL_Z1}. We calculate
separately the contributions proportional to $\tilde{w}_{12}^{(0)}$
and the strongly-ordered squared amplitude in
Eq.~\eqref{eq:integral_NLL_Z1}, starting with the former. 
Its calculation is addressed by Algorithm~\ref{algo:NLLevolution2j},
that we now outline.
We start with the truncated evolution introduced above, and starting
from the truncation scale $\kappa$ we generate two insertions of
momenta $k_a$ and $k_b$. We start by considering the two ratios of
Sudakov factors $\Delta_{12}(Q)/\Delta_{12}(k_{t(ab)})$ and
$\Delta_{12}(Q)/\Delta_{12}(k_{ta})$ in the second and third term in
the r.h.s. of Eq.~\eqref{eq:integral_NLL_Z1}, respectively.
As shown in Ref.~\cite{Banfi:2021owj}, these two terms only contribute
at NLL in the \textit{unordered} kinematic region where
$k_{ta}\sim k_{tb}^\prime\sim k_{tb}$, while they cancel in strongly
ordered regimes.
We can therefore introduce an extra ratio of Sudakov factors between
the scales $k_{ta}$ and $k_{tb}$ in the second and third terms in the
r.h.s. of Eq.~\eqref{eq:integral_NLL_Z1}, as it only amounts to
introducing subleading logarithmic corrections since~\footnote{Sec. 4
  of Ref.~\cite{Banfi:2021owj} contains a more detailed discussion
  about this point.}
\begin{equation}
\label{eq:logsApprox}
{\cal O}\left(\alpha_s\ln \frac{k_{ta}}{k_{t(ab)}}\right)\sim {\cal
 O}\left(\alpha_s\ln \frac{k_{ta}}{k_{tb}}\right) \sim {\cal
 O}\left(\alpha_s\right) \,.
\end{equation}
Concretely, for the contributions in which $k_b$ is emitted off dipole
$\{1a\}$ we can make the replacements
\begin{align}
\label{eq:extra-sudakovs}
  \abar^2(k_{t(ab)})
\frac{\Delta_{12}(Q)}{\Delta_{12}(k_{t(ab)})} &\rightarrow \abar(k_{ta})
\abar(k_{tb})\frac{\Delta_{12}(Q)}{\Delta_{12}(k_{ta})}\frac{\Delta_{1a}(k_{ta}\,k_{tb}/k_{tb}^\prime)}{\Delta_{1a}(k_{tb})}\,,\\
\abar^2(k_{ta})
\frac{\Delta_{12}(Q)}{\Delta_{12}(k_{ta})} &\rightarrow \abar(k_{ta})
\abar(k_{tb})\frac{\Delta_{12}(Q)}{\Delta_{12}(k_{ta})}\frac{\Delta_{1a}(k_{ta})}{\Delta_{1a}(k_{tb})}\,,
\end{align}
in the second and third term in the r.h.s. of
Eq.~\eqref{eq:integral_NLL_Z1}, respectively. The complementary colour
flow (i.e.\ $k_b$ is emitted off dipole $\{a2\}$) is treated
analogously.
These approximations are unnecessary from a purely theoretical point
of view (they introduce at most NNLL corrections). However, the extra
ratio of Sudakov factors has the advantage of suppressing regions of
phase space close to the collinear singularity, therefore guaranteeing
a much improved numerical stability in the calculation.
Following exactly the same reasoning, we can also replace $k_{t(ab)}$
with $k_{ta}$ in the scale of $\textit{all}$ the $Z^{(0)}$ generating
functionals in the second term in the r.h.s. of
Eq.~\eqref{eq:integral_NLL_Z1}. This also preserves the collinear
safety of the latter.

\begin{algorithm}[h]
\hspace{-0.3cm} Generate emissions $k_a$ and $k_b$ according to
Algorithms~\ref{algo:ka} and~\ref{algo:kb}, respectively\;
\hspace{-0.2cm}Create an \textit{event}:\\
  \While{{\rm true}}
  {
      \uIf{$\Theta_{\rm in} (k_a) = 1$ and $\Theta_{\rm
          in} (k_b) = 1$} {
      Reconstruct $k_{(ab)}$ with Eq.~\eqref{eq:map_ab}\;
      Fill the histogram with $V(k_{(ab)})$ iff $\Theta_{\rm in} (k_{(ab)}) = 1$ and break\;}
      \ElseIf{$\Theta_{\rm in} (k_a) = 1$ or $\Theta_{\rm
        in} (k_b) = 1$} {
        Fill the histogram with either
        $V(k_a)$ or $V(k_b)$ and break\;}
    Split the emitting dipole into three adjacent dipoles according to
    $k_a$ and $k_b$\;
    Apply Algorithm~\ref{algo:LLevolution} with starting scale
    $k_{ta}$ and evolution time~\eqref{eq:time_NLL}\;
    break\;
     }
\hspace{-0.3cm}     Create a \textit{counter-event}:\\
     Set $w \mathrel{*}= -1$\;
  \While{{\rm true}}
  {
    Reconstruct $k_{(ab)}$ with Eq.~\eqref{eq:map_ab}\;
    Replace $k_a$ with $k_{(ab)}$\;
    \If{$\Theta_{\rm in} (k_{(ab)}) = 1$} {Fill the
      histogram with $V(k_{(ab)})$ and break\;}
    Split the emitting dipole into two adjacent dipoles according to $k_{(ab)}$\;
    Apply Algorithm~\ref{algo:LLevolution} with starting scale
    $k_{ta}$ and evolution time~\eqref{eq:time_NLL}\;
    break\;
     }
\caption{Insertion of $Z^{(1)}_{12}[Q;\{u\}]$ starting from the scale $\kappa$}
\label{algo:NLLevolution2j}
\end{algorithm}
\addtocounter{algocf}{-1}
\addtocounter{subalgo}{1}
In Eq.~\eqref{eq:extra-sudakovs}, the argument of the second Sudakov
is such that we can generate
\begin{equation}
\label{eq:newbound}
k_{tb} \leq k_{ta} \frac{k_{tb}}{k_{tb}^\prime} = k_{ta}\,f(p_{\ell_1},p_{\ell_2},\hat{k}_b)\,,
\end{equation}
where we used
\begin{equation}
k_{tb}^\prime = \frac{k_{tb}}{f(p_{\ell_1},p_{\ell_2},\hat{k}_b)}\,.
\end{equation}
The function $f(p_{\ell_1},p_{\ell_2},\hat{k}_b)$, defined in
Eq.~\eqref{eq:bound}, exclusively depends on the directions of the
momenta $p_{\ell_1}$, $p_{\ell_2}$, $k_b$ and not on their
energies. Eq.~\eqref{eq:newbound} arises from requiring
$k_{tb}^\prime < k_{ta}$.
The first ($k_a$) and second ($k_b$) insertions are then generated
according to Algorithms~\ref{algo:ka} and~\ref{algo:kb}, respectively.

We then calculate the difference of terms proportional to
$\tilde{w}_{12}^{(0)}$ in Eq.~\eqref{eq:integral_NLL_Z1}.
An important remark concerns the construction of the $k_{(ab)}$
momentum appearing in the counter-term in
Eq.~\eqref{eq:integral_NLL_Z1}. This momentum is introduced in
Eq.~\eqref{eq:map_ab} where it is defined in the rest frame of the
dipole that radiates the pair $k_a$, $k_b$ (with dipole axis along the
$z$ direction), and needs to be Lorentz transformed back into the
event frame. All Lorentz transformations are performed as described in
Sec.~\ref{sec:ll-evolution}, using momenta with unit energy.
A last comment about the procedure to fill the histograms in
Algorithm~\ref{algo:NLLevolution2j} is in order. In particular, in
order to eliminate NNLL contributions, when both insertions $k_a$ and
$k_b$ are in the interjet rapidity gap in the \textit{event} we
consider the observable calculated on the massless parent defined in
Eq.~\eqref{eq:map_ab}. This procedure exactly reproduces what is done
in the \textit{counter-event}, and therefore ensures that for the NNLL
configurations in which both insertions are in the gap the two
contributions cancel by construction.

Finally, the remaining contribution to the
evolution~\eqref{eq:integral_NLL_Z1} of $Z^{(1)}_{12}[Q;\{u\}]$,
proportional to the strongly ordered double-soft squared amplitude
($w^{(0)}_{12}w^{(0)}_{a2}$ and $w^{(0)}_{12}w^{(0)}_{1a}$), is also
obtained with Algorithm~\ref{algo:NLLevolution2j}. Here one must
replace all instances of $k_{(ab)}$ with $k_a$, and generate $k_b$
with the simpler constraint $k_{tb} < k_{ta}$, which is obtained by
setting the angular function $f(p_{\ell_1},p_{\ell_2},\hat{k}_b)\to 1$
in Eq.~\eqref{eq:bound}.  Moreover, we set the weight according to the
strongly-ordered limit of Eq.~\eqref{eq:weight}
(cf. Eq.~\eqref{eq:integral_NLL_Z1}), that simply amounts to removing
the reweighing step~\eqref{eq:weight} altogether.
This guarantees that the iteration of the LL evolution kernel
in~\eqref{eq:integral_NLL_Z0} is correctly subtracted.
The final result for ${\cal H}_2\otimes S_2(v)$ is then obtained as
the sum of the result of the above three contributions.

\begin{algorithm}[htp]
  Pick the dipole that emits $k_b$ among $\{p_{\ell_1} k_a\}$ and
  $\{k_a p_{\ell_2}\}$ with probability $1/2$\;
  Update weight $w \mathrel{*}= 2$\;
  Generate $k_{tb}$ w.r.t. the emitting dipole as in Algorithm~\ref{algo:LLevolution} starting from the scale
  \begin{equation}
    \label{eq:bound}
    k_{ta} \,f(p_{\ell_1},p_{\ell_2},\hat{k}_b)\,,\quad
    f(p_{\ell_1},p_{\ell_2},\hat{k}_b)\coloneqq\sqrt{\frac{(p_{\ell_1} \cdot p_{\ell_2})}{2 (p_{\ell_1} \cdot \hat{k}_b)(\hat{k}_b \cdot p_{\ell_2})}}\,,
  \end{equation}
  where $\hat{k}_b$ denotes $k_b$ with its $k_t$ w.r.t. the emitting dipole
  set to one\;
  Update the weight
  \begin{equation}
    \label{eq:weight}
    w \mathrel{*}= \frac{\tilde{w}^{(0)}_{12}(k_b,k_a)}{w^{(0)}_{12}(k_a)
    w^{(0)}_{1a}(k_b)}\,;\quad w \mathrel{*}= \frac{\tilde{w}^{(0)}_{12}(k_a,k_b)}{w^{(0)}_{12}(k_a) w^{(0)}_{a2}(k_b)}\,,
\end{equation}
for the two dipoles, respectively\;
\caption{Generation of emission $k_{b}$ starting from the scale $k_{ta}$}
\label{algo:kb}
\end{algorithm}
%
\renewcommand{\thealgocf}{\arabic{algocf}}

\section{Numerical results for the $E_t$ distribution in the interjet
  gap at NLL}
\label{sec:results}
In this section we apply the technique described in
Sec.~\ref{sec:PT-solution} to the calculation of the transverse energy
distribution in the rapidity gap between the two cone jets in
$e^+e^-$.
In the following we set $\sqrt{s} = M_Z$ and adopt the value
$\alpha_s(M_Z) = 0.118$ for the strong coupling constant.
To obtain a physical prediction for this observable, we introduce the
standard perturbative scales used in resummed calculations whose
variation quantifies the size of subleading logarithmic corrections.
These are discussed in Appendix~\ref{sec:scales}. 
We vary the renormalisation scale $\mu_R$ by a factor of two around
its central value $\mu_R=\sqrt{s}$, and for central $\mu_R$, we also
vary the resummation scale $\mu_Q$ by a factor of two around its
central value $\mu_Q=\sqrt{s}/2$. The final perturbative uncertainty
shown in the results that follow is obtained as the envelope of the
above five predictions.
The calculation performed in this section is strictly speaking valid
only in the limit of soft radiation, and therefore should be
consistently matched to a fixed order calculation in the bulk of the
phase space where the emitted radiation is hard.
A matching of this type is standard in resummed calculation and must
be performed in future phenomenological applications.

As a check of our calculation, we have also computed the LL
$\Sigma(v)$ cumulative distribution as a function of the evolution
variable $t$ defined in Eq.~\eqref{eq:time}, and reproduced the
results of Ref.~\cite{Dasgupta:2002bw}. Notice that the LL evolution
time~\eqref{eq:time} is defined in terms of the \textit{dipole}
transverse momentum of the gluon that is radiated inside the interjet
rapidity gap. This transverse momentum is related to the physical
observable $E_t$ via an ${\cal O}(1)$ angular function that depends on
the orientation of the emitting dipole w.r.t. the thrust axis, that
varies on an event-by-event basis. Therefore the relation between $t$
and $E_t$ is not bijective.
In Ref.~\cite{Banfi:2021owj} we have also compared the
${\cal O}(\alpha_s^2)$ expansion of our calculation to fixed-order
predictions in full QCD, finding excellent agreement in the limit of
$E_t\to 0$. In the same article, we have also verified that our
fixed-order expansion for the energy distribution $E$ reproduces the
calculation of Ref.~\cite{Becher:2016mmh}.
As a further check, we have carried out two independent
implementations of the algorithms given in Sec.~\ref{sec:PT-solution}
and found complete agreement. A public version of the code can be
found in Ref.~\cite{gnole}.

\begin{figure}[htbp]
 \centering
 \includegraphics[trim={0 1cm 0 0}, clip]{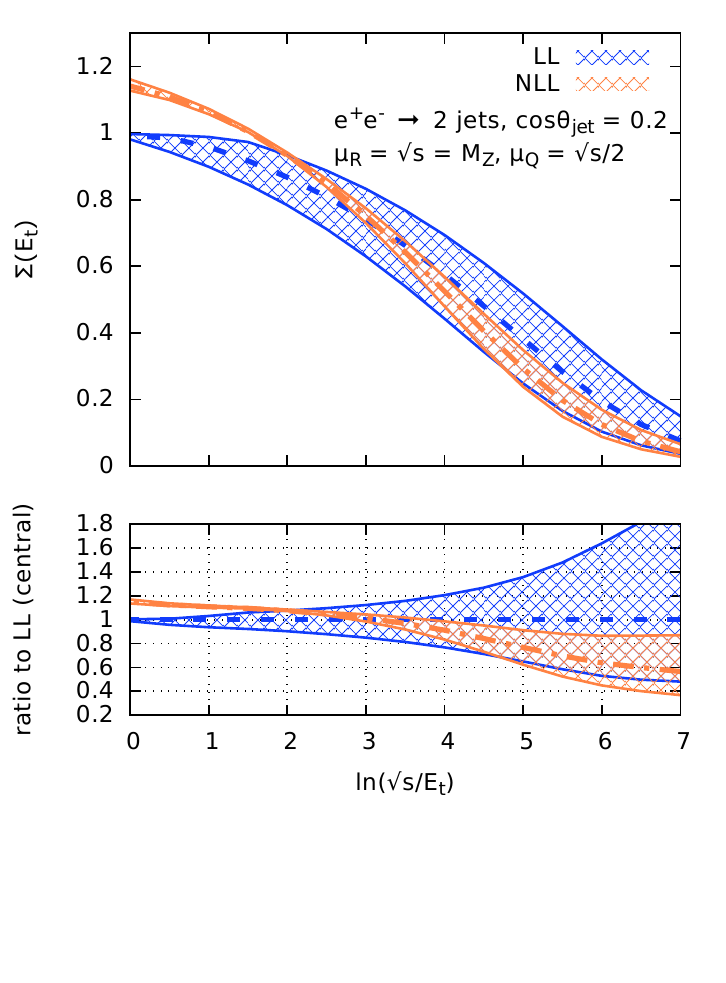}%
 \includegraphics[trim={0 1cm 0 0}, clip]{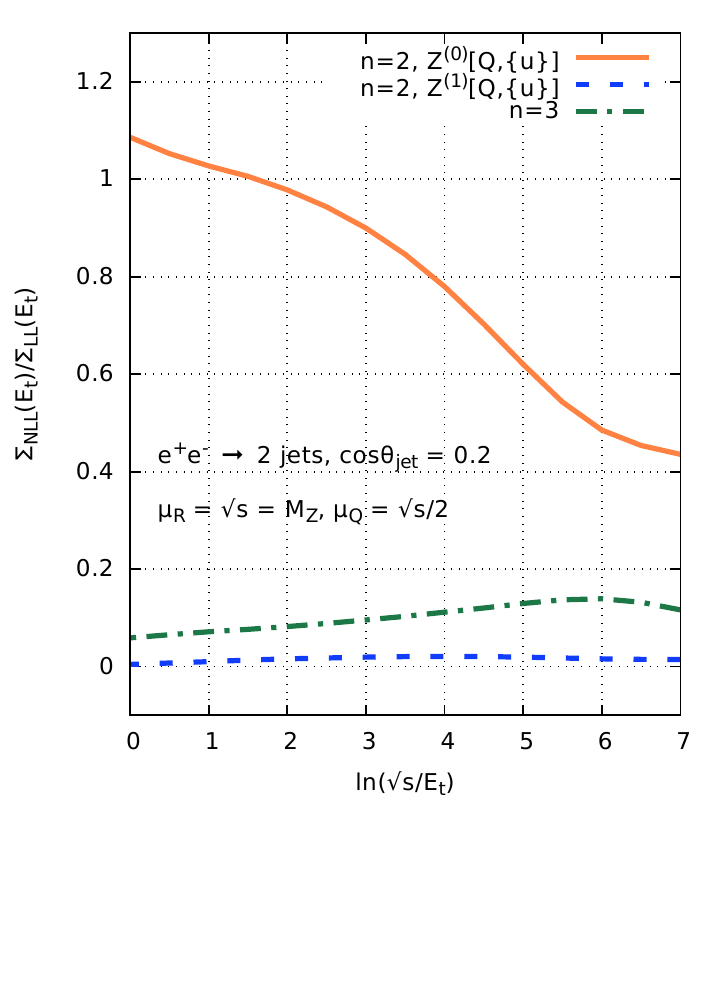}%
 \caption{Left: cumulative distribution $\Sigma(E_t)$ for the
   transverse energy in the interjet rapidity gap at LL and NLL for
   $c=0.2$. Right: breakdown of the
   contributions to the NLL correction, relative to the LL
   prediction.}
 \label{fig:results-c02}
\end{figure}

\begin{figure}[htbp]
 \centering
 \includegraphics[trim={0 1cm 0 0}, clip]{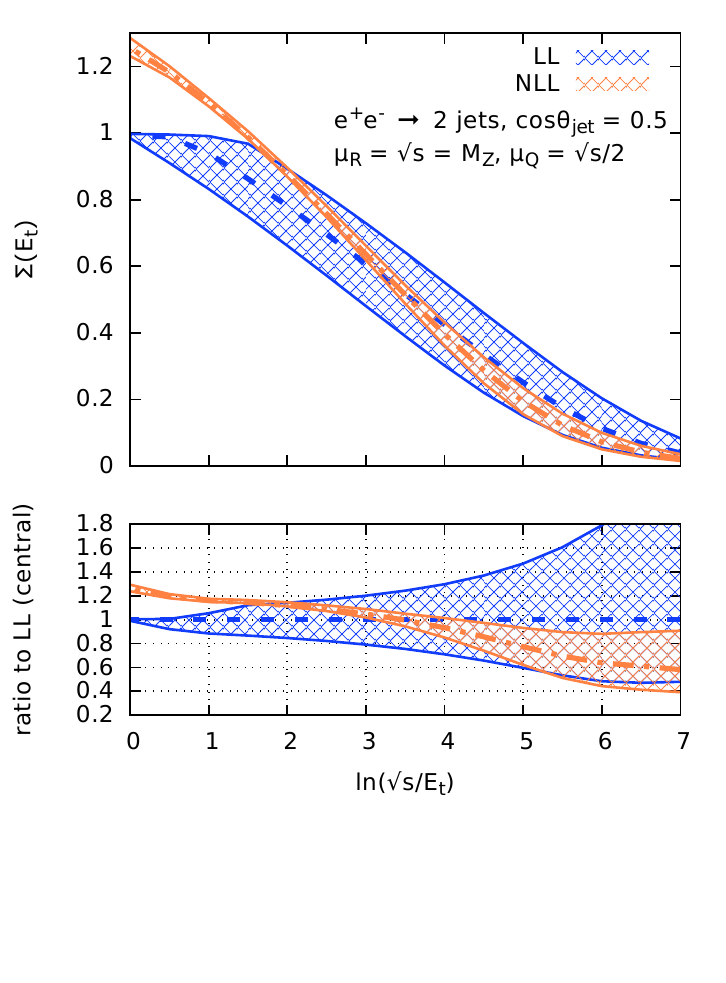}%
 \includegraphics[trim={0 1cm 0 0}, clip]{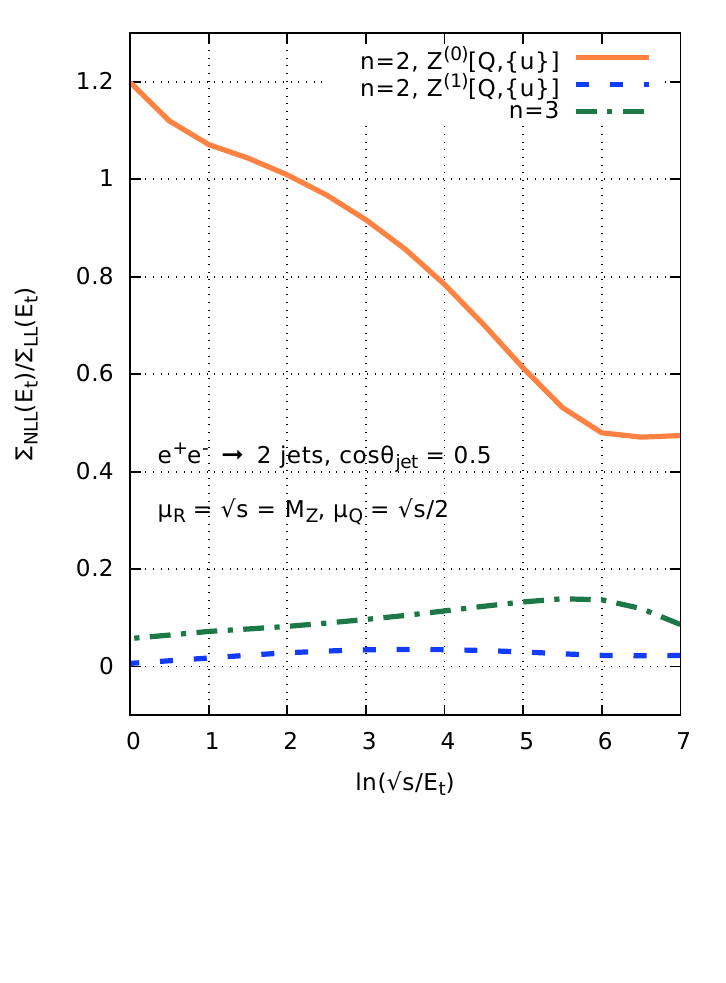}%
 \caption{Left: cumulative distribution $\Sigma(E_t)$ for the
   transverse energy in the interjet rapidity gap at LL and NLL for
   $c=0.5$. Right: breakdown of the
   contributions to the NLL correction, relative to the LL
   prediction.}
 \label{fig:results-c05}
\end{figure}

\begin{figure}[htbp]
 \centering
 \includegraphics[trim={0 1cm 0 0}, clip]{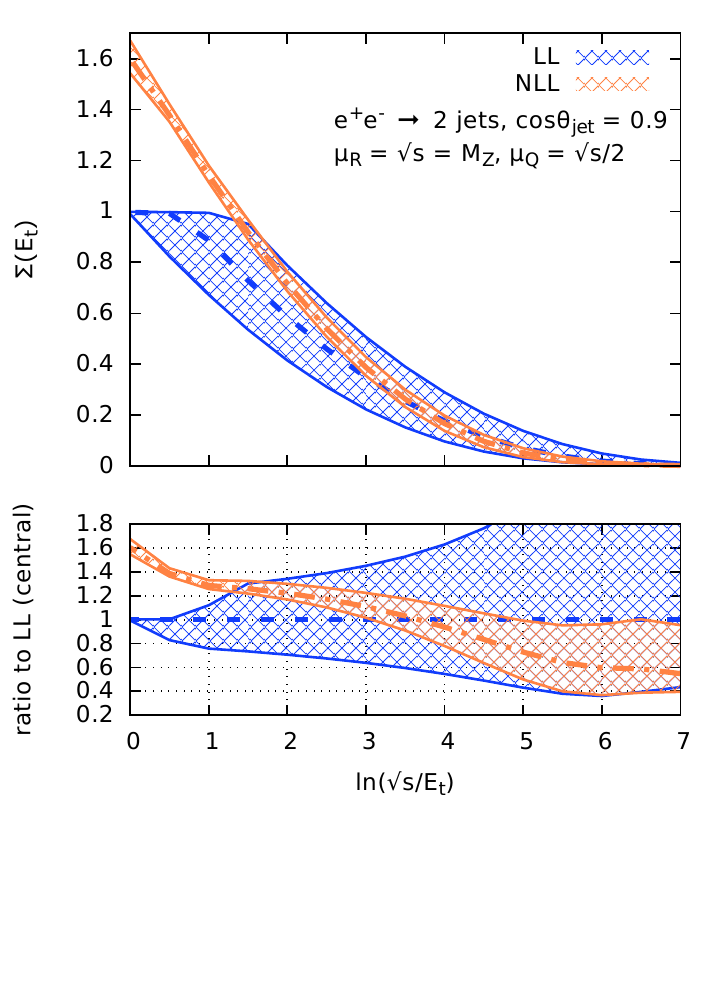}%
 \includegraphics[trim={0 1cm 0 0}, clip]{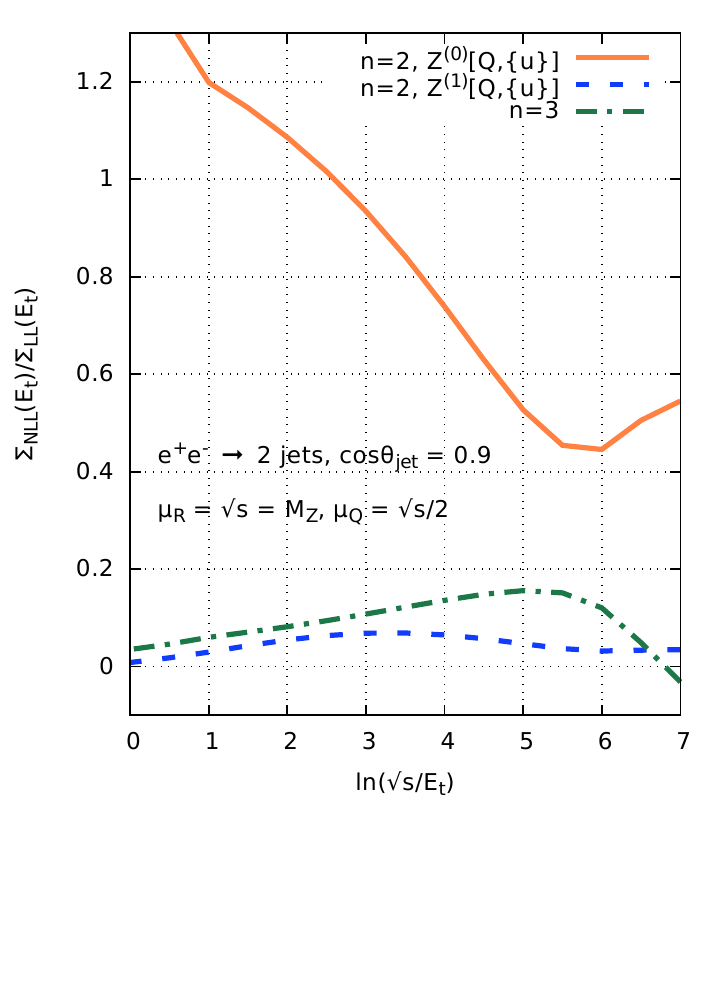}%
 \caption{Left: cumulative distribution $\Sigma(E_t)$ for the
   transverse energy in the interjet rapidity gap at LL and NLL for
   $c=0.9$. Right: breakdown of the
   contributions to the NLL correction, relative to the LL
   prediction.}
 \label{fig:results-c09}
\end{figure}

In the left plots of
Figs.~\ref{fig:results-c02},~\ref{fig:results-c05},~\ref{fig:results-c09},
we report the cumulative distribution~\eqref{eq:sigma-def} at LL and
NLL for three different values for the width of the interjet rapidity
gap ($\cos\theta_{\rm jet}=c=\{0.2, 0.5, 0.9\}$), which correspond to
different opening angles of the two hard jets
(cf. Eq.~\eqref{eq:c_def}).
We observe that the definition of the infrared scale $Q_0$ in
Eq.~\eqref{eq:Q0def} (introduced in our prescription to deal with the
Landau singularity in Eq.~\eqref{eq:initial-cond-4D-alt}) acts as a
cutoff on the transverse momentum of the emissions w.r.t. the emitting
dipole. Therefore, the region of the plots in which the observable
$E_t\sim Q_0$ (and below) is susceptible to non-perturbative
effects. We therefore truncate the plots at this scale (i.e.
$\ln\sqrt{s}/E_t=1/(2\alpha_s\beta_0)\simeq 7$) cutting out the
non-perturbative region.
We start by considering the region of the plots that corresponds to a
large transverse energy inside the interjet rapidity gap.
The predictivity here is restored upon a matching to a fixed-order
calculation, which however is not performed in
Figs.~\ref{fig:results-c02},~\ref{fig:results-c05},~\ref{fig:results-c09}. 
We therefore do not comment further on this region and we rather focus
on the small $E_t$ regime, where (non-global) resummation effects are
dominant. We notice, however, that the normalisation of the curves at
large $E_t$ changes with the size of the cone jets. This can
understood by observing the $c$-dependence of the hard factors
${\cal H}_2$ and ${\cal H}_3$ calculated in
Sec.~\ref{sec:PT-solution}. The residual scale dependence at large
$E_t$ present at NLL is due to the $\mu_R$ dependence of the hard
factors, which is absent by construction at LL.
At small $E_t$, we observe that NLL corrections are large and
negative, and reach $40\%$ in size when the resummed logarithms are
large, consistently across different values of the jet cone size. 
We stress that the results presented here adopt the physical $E_t$
definition for the LL calculation rather than its strict
leading-logarithmic limit in which $E_t$ is simply the dipole
tranverse momentum of the emission in the interjet gap. In the latter
case, the size of the genuine NLL corrections would be even larger as
these would also compensate for the kinematical difference between the
dipole $k_t$ and the actual definition of $E_t$.
We also observe a substantial reduction of the perturbative
uncertainty, up to a factor of two, in the NLL calculation compared to
the LL prediction, whose uncertainty band however accounts for the
large NLL corrections.

It is informative to study the size of the various contributions to
the NLL corrections to the cumulative distribution. In the right plots
of
Figs.~\ref{fig:results-c02},~\ref{fig:results-c05},~\ref{fig:results-c09},
we then show the breakdown of the NLL correction into three different
pieces. These are the ${\cal H}_n\otimes S_n(v)$ terms in
Eq.~\eqref{eq:master}, with $n=2$ and $n=3$. Moreover, for $n=2$, we
plot separately the contribution from the functional $Z_{12}^{(0)}$,
defined in Eq.~\eqref{eq:integral_NLL_Z0}, and $Z^{(1)}_{12}$, defined
in Eq.~\eqref{eq:integral_NLL_Z1}.
We observe that for small interjet gaps, i.e.\ fat cone jets, the NLL
result is completely dominated by the $Z_{12}^{(0)}$ correction, with
an additional sizeable correction from the ${\cal H}_3\otimes S_3(v)$
term.
For larger $c$ values (corresponding to narrower jets), however, the
contribution of the $Z^{(1)}_{12}$ correction slightly grows and
becomes comparable to that of the ${\cal H}_3\otimes S_3(v)$ term at
small values of $E_t$.
The moderate contribution of the $Z^{(1)}_{12}$ correction compared to
$Z^{(0)}_{12}$ justifies entirely the perturbative approach adopted in
Eq.~\eqref{eq:pert} for the generating functional $Z_{12}$ which led
to the evolution equations~\eqref{eq:integral_NLL_Z0}
and~\eqref{eq:integral_NLL_Z1}.

\section{Conclusions}
\label{sec:conclusions}
In this article we have presented the first NLL resummation for
non-global radiative corrections to a collider observable in the
large-$N_c$ limit.
We considered a set of evolution equations that we derived in a recent
article, describing the dynamics of soft gluons emitted at large
angles with respect to the hard scattering process.
To solve these equations numerically, we have reformulated the
resummation in terms of generating functionals, which can be used for
the calculation of the emission probability associated with any given
final state containing $n$ soft gluons.
This formulation is suitable for a numerical implementation by means
of Monte Carlo methods. Using this technology, we have presented an
algorithm for the solution of the evolution equations, hence achieving
the resummation of NLL non-global corrections.
We applied this formalism to the calculation of the NLL distribution
of the transverse energy $E_t$ of the radiation within the gap between
two hard cone jets produced in $e^+e^-$ collisions. We found that the
NLL corrections are rather sizeable and their inclusion leads to a
substantial reduction of the perturbative scale uncertainties in the
theoretical calculation. They do therefore play an important role in
the accurate prediction of this class of observables at particle
colliders.

We point out that the algorithms presented in this article are not
tailored to the specific observable considered here, and are directly
applicable to hadron-collider observables in the large-$N_c$
approximation.
This is because the only process dependence occurs in the hard factors
${\cal H}_n$ of Eq.~\eqref{eq:master} (which in turn can be calculated
algorithmically for a given process), while the evolution of the soft
factors $S_n$ with the energy scale is entirely process independent as
expected from the factorisation of squared amplitudes in the soft
limit.
Therefore, the formalism presented here can be applied to all
observables that are solely sensitive to soft radiation emitted with
large angles from the hard scattering, which are characterised by a
single logarithmic perturbative expansion, i.e.\ the dominant
logarithmic tower in the cumulative distribution is of the form
$\alpha_s^nL^n$.
The application to observables sensitive to collinear radiation, on
the other hand, requires extra care since the formulation presented
here must be supplemented with the correct resummation of the
corresponding collinear logarithms (obtained for instance with
standard techniques for global observables), and the double counting
between the two regions (i.e.\ the soft \text{and} collinear limit)
must be consistently subtracted in the evolution equations for the
generating functionals.
This subtraction is simple for most observables, and it simply amounts
to dividing the cumulative distribution~\eqref{eq:master} by the
contribution from \textit{primary} radiation.
This is obtained by running the algorithms given in
Sec.~\ref{sec:PT-solution} while forbidding the dipoles to split, so
that all emissions are radiated off the primary $q\bar{q}$ dipole.
However, we notice that there might be cases in which the subtraction
becomes conceptually more delicate, such as in the case of observables
affected by abelian \textit{clustering}
logarithms~\cite{Banfi:2005gj}.
The technical details of these subtraction procedures, as well as the
application to hadronic observables is left for future work.

The computer code \textsc{Gnole}~\cite{gnole} used to perform the
calculations presented in this article can be downloaded from the
repository:
\begin{center}
\url{https://github.com/non-global/gnole}
\end{center}

\section*{Acknowledgments}
We are grateful to Gavin Salam for several useful discussions through
the course of this work, and to Mrinal Dasgupta, Keith Hamilton, Gavin
Salam, and Gregory Soyez for numerous exchanges about conceptual
aspects of non-global resummation and its Monte Carlo implementation.
We would also like to thank Mrinal Dasgupta and Gavin Salam for
sharing with us the computer code developed in
Ref.~\cite{Dasgupta:2002bw}, that we used as a benchmark and for
comparisons in our numerical tests, as well as for their constructive
comments on the article.
This work was supported by the Science Technology and Facilities
Council (STFC) under grants number ST/T00102X/1 (AB) and ST/T000864/1
(FD), as well as by a Royal Society Research Professorship
(RP$\backslash$R1$\backslash$180112) and University Research
Fellowship (URF$\backslash$R1$\backslash$211294) (FD).

\appendix

\section{Symmetries of the squared amplitude and choice of ordering}
\label{sec:symmetries}
In this appendix we comment on the choice of the dipole $k_t$ as
ordering variable for the evolution equation.
Let us start by considering the emission of two soft gluons in the
strongly ordered limit, and we focus on the double-real integral
\begin{align}
  \label{eq:real-toy}
\frac{1}{2!}\int [d k_a] \int [d k_b] \left[w^{(0)}_{12}(k_a)
  w^{(0)}_{1a}(k_b)+w^{(0)}_{12}(k_a) w^{(0)}_{a2}(k_b)
  \right] u(k_a)u(k_b)\,.
\end{align}
We now introduce a partition of unity using the \textit{dipole}
transverse momentum
\begin{equation}
\label{eq:partition}
1=\Theta(k_{ta}-k_{tb})+\Theta(k_{tb}-k_{ta})\,.
\end{equation}
We stress that the dipole $k_{tb}$ is defined differently in the two
colour flow configurations of Eq.~\eqref{eq:real-toy}, following
Eq.~\eqref{eq:ktij-def}, which highlights the non-trivial nature of
the partition~\eqref{eq:partition}.
In order to write a Monte Carlo algorithm that iterates correctly the
squared amplitude, the two regions in Eq.~\eqref{eq:partition} must be
identical (or at least in the kinematic limits relevant to a given
logarithmic accuracy), so that the combinatorial factor $1/2!$ in
Eq.~\eqref{eq:real-toy} can be effectively replaced with a kinematic
ordering.
To show that dipole $k_t$ ordering is a suitable choice, we observe
that the squared amplitude in the strongly-ordered limit satisfies the
symmetry (e.g.\ in the case of two emissions)
\begin{align}
\label{eq:Tdef}
\mathbb{T} \coloneqq \{\hat{n}_a\leftrightarrow \hat{n}_b;
k_{tb}\leftrightarrow k_{ta}\}\,,
\end{align}
where the transverse momenta are always meant w.r.t.\ the emitting
dipole and the directions $\hat{n}_{a,b}$ are taken in the $\{12\}$
dipole frame. This symmetry is highly non trivial, and it is now
interesting to understand how the integrand in Eq.~\eqref{eq:real-toy}
behaves under its action.
We consider separately the two colour flows
\begin{align}
{\rm 1):}\quad \{1,a,b,2\}\,,\qquad
{\rm 2):}\quad \{1,b,a,2\}\,,
\end{align}
and define the transformation $\mathbb{T}^{(i)}$ ($i=1,2$) such that
for each colour flow we can obtain the transformed momenta
$\tilde{k}_{a,b}$ as
\begin{align}
\label{eq:transform}
        \begin{bmatrix}
          \tilde{k}_{a} \\
          \tilde{k}_{b}
        \end{bmatrix}
= \mathbb{T}^{(i)} 
       \begin{bmatrix}
          k_{a} \\
          k_{b}
        \end{bmatrix}\,.
\end{align}
Imposing the constraints of Eq.~\eqref{eq:Tdef} leads to
\begin{align}
\mathbb{T}^{(1)} = 
\begin{bmatrix}
0 \qquad        &  \left(\frac{(p_1 \cdot k_b) (k_a \cdot
    p_2)}{(p_1\cdot p_2)(k_a\cdot k_b)}\right)^{-1/2} \\
\left(\frac{(p_1\cdot k_a) (k_b \cdot p_2)}{(p_1\cdot p_2)(k_a\cdot k_b)}\right)^{1/2} \qquad & 0
\end{bmatrix}\,,\\
\mathbb{T}^{(2)} = 
\begin{bmatrix}
0 \qquad        &  \left(\frac{(p_1\cdot k_a) (k_b \cdot
    p_2)}{(p_1\cdot p_2)(k_a\cdot k_b)}\right)^{-1/2} \\
\left(\frac{(p_1 \cdot k_b) (k_a \cdot p_2)}{(p_1\cdot p_2)(k_a\cdot k_b)}\right)^{1/2} \qquad & 0
\end{bmatrix}\,.
\end{align}
In the above matrices $\mathbb{T}^{(i)}$ each entry is to be
understood to be proportional to the identity operator that acts on
the four momenta as in Eq.~\eqref{eq:transform}.
Let us also work out the action of the above transformation on the
phase space measure. We find
\begin{equation}
\mathbb{T}^{(1)} \left[[dk_a] [d k_b]\right] = \frac{(p_1\cdot k_a) (k_b\cdot
   p_2)}{(p_1\cdot k_b) (p_a\cdot p_2)}\,[dk_a] [d k_b] \,,\,\,\, 
\mathbb{T}^{(2)} \left[[dk_a] [d k_b]\right] = \frac{(p_1 \cdot k_b) (k_a\cdot
   p_2)}{(p_1\cdot k_a) (k_b\cdot p_2)}\,[dk_a] [d k_b] \,.
\end{equation}
The strongly ordered squared amplitude is invariant under the above
trasformations \textit{for each separate dipole}, a consequence of the
conformal symmetry of the integrand in the presence of strongly
ordered kinematics. However, due to the non-trivial effect of
$\mathbb{T}$ on the phase space measure, the LL integrand itself is
\textit{not} invariant.
By applying the transformation to the integrand we find (e.g. for
dipole $1$)
\begin{equation}
\mathbb{T}^{(1)} \left[[dk_a] [d k_b]
  w^{(0)}_{12}(k_a) w^{(0)}_{a2}(k_b)\right] =  [dk_a] [d k_b] w^{(0)}_{12}(k_a)
  w^{(0)}_{1a}(k_b) \,,
\end{equation}
that is the transformation simply maps the integrand into the one
corresponding to the complementary colour configuration. This implies
that the full LL integrand, given by the sum of the different dipoles
is indeed invariant under $\mathbb{T}$, and that therefore the
transverse momentum taken w.r.t.\ the emitting dipole can be adopted as
an evolution variable as done in Eq.~\eqref{eq:NLL-evolution-kt-diff}.

At NLL we need to consider unordered configurations in which the two
emissions $k_a$ and $k_b$ are described by the full double-soft
squared amplitude in large-$N_c$
$\tilde{w}^{(0)}_{12}(k_a,k_b)+\tilde{w}^{(0)}_{12}(k_b,k_a)$ (see
Eq.~\eqref{eq:tildew}). This squared amplitude is not invariant under
the $\mathbb{T}$ transformations, indicating that dipole $k_t$
ordering cannot be used for the calculation of the double-real
correction~\eqref{eq:Ureal}. Instead, one can order the emissions
$k_a$ and $k_b$ using their transverse momenta w.r.t.\ the $\{12\}$
dipole $k_{ta}$, $k_{tb}^\prime$, under which the squared amplitude is
fully symmetric. This explains the factor
$\Theta(k_{ta}-k_{tb}^\prime)$ in Eq.~\eqref{eq:Ureal}.
An alternative solution would be to formulate the whole evolution
ordered in \textit{energy}, which would allow one to use the same
ordering in the strongly ordered limit as well as in unordered
kinematic configurations.

\section{Dependence on the perturbative scales $\mu_R$ and $\mu_Q$}
\label{sec:scales}
In this appendix we introduce the renormalisation $\mu_R$ and
resummation $\mu_Q$ scales. In general, there are two sources of
$\mu_R$ dependence, which appears both in the hard factors
${\cal H}_2$ and ${\cal H}_3$ as well as in the soft factors $S_2$ and
$S_3$ (or equivalently in the generating functionals
$Z_{12}[Q;\{u\}]$, $Z_{13}[Q;\{u\}]$ and $Z_{23}[Q;\{u\}]$ in
Eq.~\eqref{eq:master_genfun}).

For the physical process under consideration, the $\mu_R$ dependence
in the hard factors is entirely encoded in the $\overline{\rm MS}$
coupling
\begin{equation}
  \alpha_s\to \alpha_s(\mu_R)\,.
\end{equation}
Extra dependence on $\mu_R$ in ${\cal H}_2$ and ${\cal H}_3$ arises
for processes which are mediated by QCD interactions at the Born
level, such as jet production at hadron colliders.
The second source of $\mu_R$ dependence is given by the generating
functionals. This is introduced by expressing $\abar(k_t)$ in terms of
$\abar\left(\frac{\mu_R}{\sqrt{s}}k_t\right)$ in
Eq.~\eqref{eq:integral_NLL}, and then expanding out the result in
$\abar\left(\mu_R\right)$ at \textit{fixed}
$\abar\left(\mu_R\right)\ln\frac{\sqrt{s}}{k_t}\sim 1$.
The running of the coupling must match the logarithmic order of the
calculation, and therefore we use one-loop running at LL and two-loop
running at NLL.

The resummation scale $\mu_Q$ is introduced to estimate the size of
subleading logarithmic corrections. Its dependence is entirely encoded
in the soft factors, and thus in the generating
functional.
The whole $\mu_R$ and $\mu_Q$ scale dependence can be easily encoded in the
evolution algorithms presented in
Sec.~\ref{sec:functionals}. Specifically, it amounts to replacing the
evolution times~\eqref{eq:time}~\eqref{eq:time_NLL} with
\begin{align}
\label{eq:time_scales_LL}
t \rightarrow & \,\tilde{t} \coloneqq -
\frac{N_c}{2\pi\beta_0}\ln \left(1-2 \tilde{\lambda}\right)\,,\qquad
                \tilde{\lambda}=\beta_0\alpha_s(\mu_R)\ln\frac{\mu_Q}{k_t}\,,
\end{align}
\begin{align}
\label{eq:time_scales_NLL}
t \rightarrow & \,\tilde{t} \coloneqq -
    \frac{N_c}{2\pi\beta_0}\ln \left(1-2\tilde{\lambda}\right) + \abar(\mu_R) \frac{\tilde{\lambda}}{1-2\tilde{\lambda}}\ln\frac{\mu_R^2}{\mu_Q^2}+\abar(\mu_R) \ln\frac{\sqrt{s}}{\mu_Q}\notag\\
  & +
    \abar(\mu_R)\left[\frac{\tilde{\lambda}}{1-2\tilde{\lambda}}\left(\frac{K^{(1)}}{2\pi\beta_0}-\frac{\beta_1}{\beta_0^2}\right)-\frac{\ln(1-2\tilde{\lambda})}{1-2\tilde{\lambda}}\frac{\beta_1}{2\beta_0^2}\right]
    + {\cal O}({\rm NNLL})\,,
\end{align}
at LL and NLL, respectively. The definition of the infrared scale
$Q_0$ given in Eq.~\eqref{eq:Q0def} is also consistently modified as
follows
\begin{equation}
\label{eq:Q0def_scales}
2\,\beta_0\alpha_s(\mu_R)\ln\frac{\mu_Q}{Q_0} = 1\,.
\end{equation}

\bibliographystyle{JHEP} \bibliography{ngl}

\providecommand{\href}[2]{#2}\begingroup\raggedright\begin{thebibliography}{10}

\bibitem{Dasgupta:2001sh}
M.~Dasgupta and G.~P. Salam, \emph{{Resummation of nonglobal QCD observables}},
  \href{http://dx.doi.org/10.1016/S0370-2693(01)00725-0}{\emph{Phys. Lett. B}
  {\bf 512} (2001) 323--330}, [\href{http://arxiv.org/abs/hep-ph/0104277}{{\tt
  hep-ph/0104277}}].

\bibitem{Dasgupta:2002bw}
M.~Dasgupta and G.~P. Salam, \emph{{Accounting for coherence in interjet E(t)
  flow: A Case study}},
  \href{http://dx.doi.org/10.1088/1126-6708/2002/03/017}{\emph{JHEP} {\bf 03}
  (2002) 017}, [\href{http://arxiv.org/abs/hep-ph/0203009}{{\tt
  hep-ph/0203009}}].

\bibitem{Banfi:2002hw}
A.~Banfi, G.~Marchesini and G.~Smye, \emph{{Away from jet energy flow}},
  \href{http://dx.doi.org/10.1088/1126-6708/2002/08/006}{\emph{JHEP} {\bf 08}
  (2002) 006}, [\href{http://arxiv.org/abs/hep-ph/0206076}{{\tt
  hep-ph/0206076}}].

\bibitem{Hatta:2013iba}
Y.~Hatta and T.~Ueda, \emph{{Resummation of non-global logarithms at finite
  $N_c$}}, \href{http://dx.doi.org/10.1016/j.nuclphysb.2013.06.021}{\emph{Nucl.
  Phys. B} {\bf 874} (2013) 808--820},
  [\href{http://arxiv.org/abs/1304.6930}{{\tt 1304.6930}}].

\bibitem{Hagiwara:2015bia}
Y.~Hagiwara, Y.~Hatta and T.~Ueda, \emph{{Hemisphere jet mass distribution at
  finite $N_c$}},
  \href{http://dx.doi.org/10.1016/j.physletb.2016.03.028}{\emph{Phys. Lett. B}
  {\bf 756} (2016) 254--258}, [\href{http://arxiv.org/abs/1507.07641}{{\tt
  1507.07641}}].

\bibitem{Hatta:2020wre}
Y.~Hatta and T.~Ueda, \emph{{Non-global logarithms in hadron collisions at
  $N_c$ = 3}},
  \href{http://dx.doi.org/10.1016/j.nuclphysb.2020.115273}{\emph{Nucl. Phys. B}
  {\bf 962} (2021) 115273}, [\href{http://arxiv.org/abs/2011.04154}{{\tt
  2011.04154}}].

\bibitem{Forshaw:2006fk}
J.~R. Forshaw, A.~Kyrieleis and M.~H. Seymour, \emph{{Super-leading logarithms
  in non-global observables in QCD}},
  \href{http://dx.doi.org/10.1088/1126-6708/2006/08/059}{\emph{JHEP} {\bf 08}
  (2006) 059}, [\href{http://arxiv.org/abs/hep-ph/0604094}{{\tt
  hep-ph/0604094}}].

\bibitem{Forshaw:2008cq}
J.~R. Forshaw, A.~Kyrieleis and M.~H. Seymour, \emph{{Super-leading logarithms
  in non-global observables in QCD: Colour basis independent calculation}},
  \href{http://dx.doi.org/10.1088/1126-6708/2008/09/128}{\emph{JHEP} {\bf 09}
  (2008) 128}, [\href{http://arxiv.org/abs/0808.1269}{{\tt 0808.1269}}].

\bibitem{Becher:2021zkk}
T.~Becher, M.~Neubert and D.~Y. Shao, \emph{{Resummation of Super-Leading
  Logarithms}},  \href{http://arxiv.org/abs/2107.01212}{{\tt 2107.01212}}.

\bibitem{Becher:2015hka}
T.~Becher, M.~Neubert, L.~Rothen and D.~Y. Shao, \emph{{Effective Field Theory
  for Jet Processes}},
  \href{http://dx.doi.org/10.1103/PhysRevLett.116.192001}{\emph{Phys. Rev.
  Lett.} {\bf 116} (2016) 192001}, [\href{http://arxiv.org/abs/1508.06645}{{\tt
  1508.06645}}].

\bibitem{Becher:2016mmh}
T.~Becher, M.~Neubert, L.~Rothen and D.~Y. Shao, \emph{{Factorization and
  Resummation for Jet Processes}},
  \href{http://dx.doi.org/10.1007/JHEP11(2016)019}{\emph{JHEP} {\bf 11} (2016)
  019}, [\href{http://arxiv.org/abs/1605.02737}{{\tt 1605.02737}}].

\bibitem{Caron-Huot:2015bja}
S.~Caron-Huot, \emph{{Resummation of non-global logarithms and the BFKL
  equation}}, \href{http://dx.doi.org/10.1007/JHEP03(2018)036}{\emph{JHEP} {\bf
  03} (2018) 036}, [\href{http://arxiv.org/abs/1501.03754}{{\tt 1501.03754}}].

\bibitem{Larkoski:2015zka}
A.~J. Larkoski, I.~Moult and D.~Neill, \emph{{Non-Global Logarithms,
  Factorization, and the Soft Substructure of Jets}},
  \href{http://dx.doi.org/10.1007/JHEP09(2015)143}{\emph{JHEP} {\bf 09} (2015)
  143}, [\href{http://arxiv.org/abs/1501.04596}{{\tt 1501.04596}}].

\bibitem{Banfi:2021owj}
A.~Banfi, F.~A. Dreyer and P.~F. Monni, \emph{{Next-to-leading non-global
  logarithms in QCD}},  \href{http://arxiv.org/abs/2104.06416}{{\tt
  2104.06416}}.

\bibitem{Forshaw:2009fz}
J.~Forshaw, J.~Keates and S.~Marzani, \emph{{Jet vetoing at the LHC}},
  \href{http://dx.doi.org/10.1088/1126-6708/2009/07/023}{\emph{JHEP} {\bf 07}
  (2009) 023}, [\href{http://arxiv.org/abs/0905.1350}{{\tt 0905.1350}}].

\bibitem{Rubin:2010fc}
M.~Rubin, \emph{{Non-Global Logarithms in Filtered Jet Algorithms}},
  \href{http://dx.doi.org/10.1007/JHEP05(2010)005}{\emph{JHEP} {\bf 05} (2010)
  005}, [\href{http://arxiv.org/abs/1002.4557}{{\tt 1002.4557}}].

\bibitem{Banfi:2010pa}
A.~Banfi, M.~Dasgupta, K.~Khelifa-Kerfa and S.~Marzani, \emph{{Non-global
  logarithms and jet algorithms in high-pT jet shapes}},
  \href{http://dx.doi.org/10.1007/JHEP08(2010)064}{\emph{JHEP} {\bf 08} (2010)
  064}, [\href{http://arxiv.org/abs/1004.3483}{{\tt 1004.3483}}].

\bibitem{DuranDelgado:2011tp}
R.~M. Duran~Delgado, J.~R. Forshaw, S.~Marzani and M.~H. Seymour, \emph{{The
  dijet cross section with a jet veto}},
  \href{http://dx.doi.org/10.1007/JHEP08(2011)157}{\emph{JHEP} {\bf 08} (2011)
  157}, [\href{http://arxiv.org/abs/1107.2084}{{\tt 1107.2084}}].

\bibitem{Dasgupta:2012hg}
M.~Dasgupta, K.~Khelifa-Kerfa, S.~Marzani and M.~Spannowsky, \emph{{On jet mass
  distributions in Z+jet and dijet processes at the LHC}},
  \href{http://dx.doi.org/10.1007/JHEP10(2012)126}{\emph{JHEP} {\bf 10} (2012)
  126}, [\href{http://arxiv.org/abs/1207.1640}{{\tt 1207.1640}}].

\bibitem{Schwartz:2014wha}
M.~D. Schwartz and H.~X. Zhu, \emph{{Nonglobal logarithms at three loops, four
  loops, five loops, and beyond}},
  \href{http://dx.doi.org/10.1103/PhysRevD.90.065004}{\emph{Phys. Rev. D} {\bf
  90} (2014) 065004}, [\href{http://arxiv.org/abs/1403.4949}{{\tt 1403.4949}}].

\bibitem{Becher:2016omr}
T.~Becher, B.~D. Pecjak and D.~Y. Shao, \emph{{Factorization for the light-jet
  mass and hemisphere soft function}},
  \href{http://dx.doi.org/10.1007/JHEP12(2016)018}{\emph{JHEP} {\bf 12} (2016)
  018}, [\href{http://arxiv.org/abs/1610.01608}{{\tt 1610.01608}}].

\bibitem{Neill:2016stq}
D.~Neill, \emph{{The Asymptotic Form of Non-Global Logarithms, Black Disc
  Saturation, and Gluonic Deserts}},
  \href{http://dx.doi.org/10.1007/JHEP01(2017)109}{\emph{JHEP} {\bf 01} (2017)
  109}, [\href{http://arxiv.org/abs/1610.02031}{{\tt 1610.02031}}].

\bibitem{Caron-Huot:2016tzz}
S.~Caron-Huot and M.~Herranen, \emph{{High-energy evolution to three loops}},
  \href{http://dx.doi.org/10.1007/JHEP02(2018)058}{\emph{JHEP} {\bf 02} (2018)
  058}, [\href{http://arxiv.org/abs/1604.07417}{{\tt 1604.07417}}].

\bibitem{Larkoski:2016zzc}
A.~J. Larkoski, I.~Moult and D.~Neill, \emph{{The Analytic Structure of
  Non-Global Logarithms: Convergence of the Dressed Gluon Expansion}},
  \href{http://dx.doi.org/10.1007/JHEP11(2016)089}{\emph{JHEP} {\bf 11} (2016)
  089}, [\href{http://arxiv.org/abs/1609.04011}{{\tt 1609.04011}}].

\bibitem{Becher:2017nof}
T.~Becher, R.~Rahn and D.~Y. Shao, \emph{{Non-global and rapidity logarithms in
  narrow jet broadening}},
  \href{http://dx.doi.org/10.1007/JHEP10(2017)030}{\emph{JHEP} {\bf 10} (2017)
  030}, [\href{http://arxiv.org/abs/1708.04516}{{\tt 1708.04516}}].

\bibitem{Martinez:2018ffw}
R.~\'Angeles~Mart\'\i{}nez, M.~De~Angelis, J.~R. Forshaw, S.~Pl\"atzer and
  M.~H. Seymour, \emph{{Soft gluon evolution and non-global logarithms}},
  \href{http://dx.doi.org/10.1007/JHEP05(2018)044}{\emph{JHEP} {\bf 05} (2018)
  044}, [\href{http://arxiv.org/abs/1802.08531}{{\tt 1802.08531}}].

\bibitem{Balsiger:2018ezi}
M.~Balsiger, T.~Becher and D.~Y. Shao, \emph{{Non-global logarithms in jet and
  isolation cone cross sections}},
  \href{http://dx.doi.org/10.1007/JHEP08(2018)104}{\emph{JHEP} {\bf 08} (2018)
  104}, [\href{http://arxiv.org/abs/1803.07045}{{\tt 1803.07045}}].

\bibitem{Neill:2018yet}
D.~Neill, \emph{{Non-Global and Clustering Effects for Groomed Multi-Prong Jet
  Shapes}}, \href{http://dx.doi.org/10.1007/JHEP02(2019)114}{\emph{JHEP} {\bf
  02} (2019) 114}, [\href{http://arxiv.org/abs/1808.04897}{{\tt 1808.04897}}].

\bibitem{Balsiger:2019tne}
M.~Balsiger, T.~Becher and D.~Y. Shao, \emph{{NLL${'}$ resummation of jet
  mass}}, \href{http://dx.doi.org/10.1007/JHEP04(2019)020}{\emph{JHEP} {\bf 04}
  (2019) 020}, [\href{http://arxiv.org/abs/1901.09038}{{\tt 1901.09038}}].

\bibitem{Balsiger:2020ogy}
M.~Balsiger, T.~Becher and A.~Ferroglia, \emph{{Resummation of non-global
  logarithms in cross sections with massive particles}},
  \href{http://dx.doi.org/10.1007/JHEP09(2020)029}{\emph{JHEP} {\bf 09} (2020)
  029}, [\href{http://arxiv.org/abs/2006.00014}{{\tt 2006.00014}}].

\bibitem{Ziani:2021dxr}
N.~Ziani, K.~Khelifa-Kerfa and Y.~Delenda, \emph{{Jet mass distribution in
  Higgs/vector boson + jet events at hadron colliders with $k_t$ clustering}},
  \href{http://dx.doi.org/10.1140/epjc/s10052-021-09379-z}{\emph{Eur. Phys. J.
  C} {\bf 81} (2021) 570}, [\href{http://arxiv.org/abs/2104.11060}{{\tt
  2104.11060}}].

\bibitem{Weigert:2003mm}
H.~Weigert, \emph{{Nonglobal jet evolution at finite N(c)}},
  \href{http://dx.doi.org/10.1016/j.nuclphysb.2004.03.002}{\emph{Nucl. Phys. B}
  {\bf 685} (2004) 321--350}, [\href{http://arxiv.org/abs/hep-ph/0312050}{{\tt
  hep-ph/0312050}}].

\bibitem{Hatta:2008st}
Y.~Hatta, \emph{{Relating e+ e- annihilation to high energy scattering at weak
  and strong coupling}},
  \href{http://dx.doi.org/10.1088/1126-6708/2008/11/057}{\emph{JHEP} {\bf 11}
  (2008) 057}, [\href{http://arxiv.org/abs/0810.0889}{{\tt 0810.0889}}].

\bibitem{Dasgupta:2018nvj}
M.~Dasgupta, F.~A. Dreyer, K.~Hamilton, P.~F. Monni and G.~P. Salam,
  \emph{{Logarithmic accuracy of parton showers: a fixed-order study}},
  \href{http://dx.doi.org/10.1007/JHEP09(2018)033}{\emph{JHEP} {\bf 09} (2018)
  033}, [\href{http://arxiv.org/abs/1805.09327}{{\tt 1805.09327}}].

\bibitem{Bewick:2019rbu}
G.~Bewick, S.~Ferrario~Ravasio, P.~Richardson and M.~H. Seymour,
  \emph{{Logarithmic accuracy of angular-ordered parton showers}},
  \href{http://dx.doi.org/10.1007/JHEP04(2020)019}{\emph{JHEP} {\bf 04} (2020)
  019}, [\href{http://arxiv.org/abs/1904.11866}{{\tt 1904.11866}}].

\bibitem{Dasgupta:2020fwr}
M.~Dasgupta, F.~A. Dreyer, K.~Hamilton, P.~F. Monni, G.~P. Salam and G.~Soyez,
  \emph{{Parton showers beyond leading logarithmic accuracy}},
  \href{http://dx.doi.org/10.1103/PhysRevLett.125.052002}{\emph{Phys. Rev.
  Lett.} {\bf 125} (2020) 052002}, [\href{http://arxiv.org/abs/2002.11114}{{\tt
  2002.11114}}].

\bibitem{Forshaw:2020wrq}
J.~R. Forshaw, J.~Holguin and S.~Pl\"atzer, \emph{{Building a consistent parton
  shower}}, \href{http://dx.doi.org/10.1007/JHEP09(2020)014}{\emph{JHEP} {\bf
  09} (2020) 014}, [\href{http://arxiv.org/abs/2003.06400}{{\tt 2003.06400}}].

\bibitem{Platzer:2020lbr}
S.~Pl\"atzer and I.~Ruffa, \emph{{Towards Colour Flow Evolution at Two Loops}},
   \href{http://arxiv.org/abs/2012.15215}{{\tt 2012.15215}}.

\bibitem{Hamilton:2020rcu}
K.~Hamilton, R.~Medves, G.~P. Salam, L.~Scyboz and G.~Soyez, \emph{{Colour and
  logarithmic accuracy in final-state parton showers}},
  \href{http://arxiv.org/abs/2011.10054}{{\tt 2011.10054}}.

\bibitem{Nagy:2020rmk}
Z.~Nagy and D.~E. Soper, \emph{{Summations of large logarithms by parton
  showers}},  \href{http://arxiv.org/abs/2011.04773}{{\tt 2011.04773}}.

\bibitem{Nagy:2020dvz}
Z.~Nagy and D.~E. Soper, \emph{{Summations by parton showers of large
  logarithms in electron-positron annihilation}},
  \href{http://arxiv.org/abs/2011.04777}{{\tt 2011.04777}}.

\bibitem{Karlberg:2021kwr}
A.~Karlberg, G.~P. Salam, L.~Scyboz and R.~Verheyen, \emph{{Spin correlations
  in final-state parton showers and jet observables}},
  \href{http://arxiv.org/abs/2103.16526}{{\tt 2103.16526}}.

\bibitem{Dulat:2018vuy}
F.~Dulat, S.~H\"oche and S.~Prestel, \emph{{Leading-Color Fully Differential
  Two-Loop Soft Corrections to QCD Dipole Showers}},
  \href{http://dx.doi.org/10.1103/PhysRevD.98.074013}{\emph{Phys. Rev. D} {\bf
  98} (2018) 074013}, [\href{http://arxiv.org/abs/1805.03757}{{\tt
  1805.03757}}].

\bibitem{Gellersen:2021eci}
L.~Gellersen, S.~H\"oche and S.~Prestel, \emph{{Disentangling soft and
  collinear effects in QCD parton showers}},
  \href{http://arxiv.org/abs/2110.05964}{{\tt 2110.05964}}.

\bibitem{Hamilton:2021dyz}
K.~Hamilton, A.~Karlberg, G.~P. Salam, L.~Scyboz and R.~Verheyen, \emph{{Soft
  spin correlations in final-state parton showers}},
  \href{http://arxiv.org/abs/2111.01161}{{\tt 2111.01161}}.

\bibitem{Konishi:1979cb}
K.~Konishi, A.~Ukawa and G.~Veneziano, \emph{{Jet Calculus: A Simple Algorithm
  for Resolving QCD Jets}},
  \href{http://dx.doi.org/10.1016/0550-3213(79)90053-1}{\emph{Nucl. Phys. B}
  {\bf 157} (1979) 45--107}.

\bibitem{Bassetto:1984ik}
A.~Bassetto, M.~Ciafaloni and G.~Marchesini, \emph{{Jet Structure and Infrared
  Sensitive Quantities in Perturbative QCD}},
  \href{http://dx.doi.org/10.1016/0370-1573(83)90083-2}{\emph{Phys. Rept.} {\bf
  100} (1983) 201--272}.

\bibitem{Dokshitzer:1991wu}
Y.~L. Dokshitzer, V.~A. Khoze, A.~H. Mueller and S.~I. Troian, \emph{{Basics of
  perturbative QCD}}.
\newblock 1991.

\bibitem{Campbell:1997hg}
J.~M. Campbell and E.~W.~N. Glover, \emph{{Double unresolved approximations to
  multiparton scattering amplitudes}},
  \href{http://dx.doi.org/10.1016/S0550-3213(98)00295-8}{\emph{Nucl. Phys. B}
  {\bf 527} (1998) 264--288}, [\href{http://arxiv.org/abs/hep-ph/9710255}{{\tt
  hep-ph/9710255}}].

\bibitem{GehrmannDeRidder:2005cm}
A.~Gehrmann-De~Ridder, T.~Gehrmann and E.~Glover, \emph{{Antenna subtraction at
  NNLO}}, \href{http://dx.doi.org/10.1088/1126-6708/2005/09/056}{\emph{JHEP}
  {\bf 09} (2005) 056}, [\href{http://arxiv.org/abs/hep-ph/0505111}{{\tt
  hep-ph/0505111}}].

\bibitem{Banfi:2012jm}
A.~Banfi, P.~F. Monni, G.~P. Salam and G.~Zanderighi, \emph{{Higgs and Z-boson
  production with a jet veto}},
  \href{http://dx.doi.org/10.1103/PhysRevLett.109.202001}{\emph{Phys. Rev.
  Lett.} {\bf 109} (2012) 202001}, [\href{http://arxiv.org/abs/1206.4998}{{\tt
  1206.4998}}].

\bibitem{Banfi:2014sua}
A.~Banfi, H.~McAslan, P.~F. Monni and G.~Zanderighi, \emph{{A general method
  for the resummation of event-shape distributions in $e^+e^-$ annihilation}},
  \href{http://dx.doi.org/10.1007/JHEP05(2015)102}{\emph{JHEP} {\bf 05} (2015)
  102}, [\href{http://arxiv.org/abs/1412.2126}{{\tt 1412.2126}}].

\bibitem{Banfi:2016zlc}
A.~Banfi, H.~McAslan, P.~F. Monni and G.~Zanderighi, \emph{{The two-jet rate in
  $e^+e^-$ at next-to-next-to-leading-logarithmic order}},
  \href{http://dx.doi.org/10.1103/PhysRevLett.117.172001}{\emph{Phys. Rev.
  Lett.} {\bf 117} (2016) 172001}, [\href{http://arxiv.org/abs/1607.03111}{{\tt
  1607.03111}}].

\bibitem{Banfi:2018mcq}
A.~Banfi, B.~K. El-Menoufi and P.~F. Monni, \emph{{The Sudakov radiator for jet
  observables and the soft physical coupling}},
  \href{http://dx.doi.org/10.1007/JHEP01(2019)083}{\emph{JHEP} {\bf 01} (2019)
  083}, [\href{http://arxiv.org/abs/1807.11487}{{\tt 1807.11487}}].

\bibitem{Monni:2019yyr}
P.~F. Monni, L.~Rottoli and P.~Torrielli, \emph{{Higgs transverse momentum with
  a jet veto: a double-differential resummation}},
  \href{http://dx.doi.org/10.1103/PhysRevLett.124.252001}{\emph{Phys. Rev.
  Lett.} {\bf 124} (2020) 252001}, [\href{http://arxiv.org/abs/1909.04704}{{\tt
  1909.04704}}].

\bibitem{Gustafson:1987rq}
G.~Gustafson and U.~Pettersson, \emph{{Dipole Formulation of QCD Cascades}},
  \href{http://dx.doi.org/10.1016/0550-3213(88)90441-5}{\emph{Nucl. Phys. B}
  {\bf 306} (1988) 746--758}.

\bibitem{Lonnblad:1992tz}
L.~Lonnblad, \emph{{ARIADNE version 4: A Program for simulation of QCD cascades
  implementing the color dipole model}},
  \href{http://dx.doi.org/10.1016/0010-4655(92)90068-A}{\emph{Comput. Phys.
  Commun.} {\bf 71} (1992) 15--31}.

\bibitem{Sjostrand:2004ef}
T.~Sjostrand and P.~Z. Skands, \emph{{Transverse-momentum-ordered showers and
  interleaved multiple interactions}},
  \href{http://dx.doi.org/10.1140/epjc/s2004-02084-y}{\emph{Eur. Phys. J. C}
  {\bf 39} (2005) 129--154}, [\href{http://arxiv.org/abs/hep-ph/0408302}{{\tt
  hep-ph/0408302}}].

\bibitem{Giele:2007di}
W.~T. Giele, D.~A. Kosower and P.~Z. Skands, \emph{{A simple shower and
  matching algorithm}},
  \href{http://dx.doi.org/10.1103/PhysRevD.78.014026}{\emph{Phys. Rev. D} {\bf
  78} (2008) 014026}, [\href{http://arxiv.org/abs/0707.3652}{{\tt 0707.3652}}].

\bibitem{Hoche:2015sya}
S.~H\"oche and S.~Prestel, \emph{{The midpoint between dipole and parton
  showers}}, \href{http://dx.doi.org/10.1140/epjc/s10052-015-3684-2}{\emph{Eur.
  Phys. J. C} {\bf 75} (2015) 461},
  [\href{http://arxiv.org/abs/1506.05057}{{\tt 1506.05057}}].

\bibitem{Cacciari:2015jma}
M.~Cacciari, F.~A. Dreyer, A.~Karlberg, G.~P. Salam and G.~Zanderighi,
  \emph{{Fully Differential Vector-Boson-Fusion Higgs Production at
  Next-to-Next-to-Leading Order}},
  \href{http://dx.doi.org/10.1103/PhysRevLett.115.082002}{\emph{Phys. Rev.
  Lett.} {\bf 115} (2015) 082002}, [\href{http://arxiv.org/abs/1506.02660}{{\tt
  1506.02660}}].

\bibitem{gnole}
A.~Banfi, F.~Dreyer and P.~Monni, ``{\textsc gnole}.''
  \url{https://doi.org/10.5281/zenodo.5637033}, Nov., 2021.
\newblock 10.5281/zenodo.5637033.

\bibitem{Banfi:2005gj}
A.~Banfi and M.~Dasgupta, \emph{{Problems in resumming interjet energy flows
  with $k_t$ clustering}},
  \href{http://dx.doi.org/10.1016/j.physletb.2005.08.125}{\emph{Phys. Lett. B}
  {\bf 628} (2005) 49--56}, [\href{http://arxiv.org/abs/hep-ph/0508159}{{\tt
  hep-ph/0508159}}].

\end{thebibliography}\endgroup
\end{document}